\documentclass[useAMS,usenatbib]{mn2e}

\usepackage{amsmath}
\usepackage{amssymb}
\usepackage{epsfig}
\usepackage{graphicx}
\usepackage{ifthen}
\usepackage{latexsym}
\usepackage{rotating}
\usepackage{subfigure}
\usepackage{times,epsf}
\usepackage{txfonts}
\usepackage{varioref}
\usepackage{verbatim}
\usepackage{url}
\usepackage{tikz}
\usepackage{epstopdf}
\usepackage{color}
\usepackage{array}
\usepackage{multirow}
\usepackage[T1]{fontenc}
\newcolumntype{M}{>{$\vcenter\bgroup\hbox\bgroup}c<{\egroup\egroup$}}

\newcommand{\be}{\begin{equation}}
\newcommand{\ee}{\end{equation}}
\newcommand{\bea}{\begin{eqnarray}}
\newcommand{\eea}{\end{eqnarray}}

\def\gsim{ \lower .75ex \hbox{$\sim$} \llap{\raise .27ex \hbox{$>$}} }
\def\lsim{ \lower .75ex\hbox{$\sim$} \llap{\raise .27ex \hbox{$<$}} }

\title[Oscillations of tori]
{Oscillations of radiation pressure supported tori near black holes}

\author[G.P. Mazur, O. Zanotti, A. S\k{a}dowski, B. Mishra,
W. Kluzniak] {Grzegorz P. Mazur$^{1,2}$\thanks{E-mail:
    gmazur@ifpan.edu.pl (GM)}, Olindo Zanotti$^{3}$\thanks{E-mail:
    olindo.zanotti@unitn.it (OZ)}, Aleksander
  S\k{a}dowski$^{4,5}$\thanks{E-mail: asadowsk@mit.edu (AS)},\newauthor Bhupendra Mishra$^{2}$,Wlodek Klu\'{z}niak$^{2}$\\
$^1$ Institute of Physics, Polish Academy of Sciences, al. Lotnik\'{o}w 32/46, 02-668 Warsaw, Poland \\
$^2$ Nicolaus Copernicus Astronomical Center, Bartycka 18, 00-716
Warsaw, Poland \\
$^3$ Laboratory of Applied Mathematics, University of Trento, Via
Mesiano 77, 38123 Trento, Italy \\ 
$^4$ MIT Kavli Institute for Astrophysics and Space Research,
77 Massachusetts Ave, Cambridge, MA 02139, USA\\
$^5$ Einstein Fellow}

\begin{document}

\maketitle

\label{firstpage}

\begin{abstract}
  We study the dynamics of radiation pressure supported 
  tori around Schwarzschild black holes, focusing on their oscillatory
  response to an external perturbation.  Using \texttt{KORAL}, a general relativistic radiation hydrodynamics code capable of
  modeling all radiative regimes from the optically thick to the optically thin,
  we monitor a sample of models at different initial temperatures and
  opacities, evolving them in two spatial dimensions for $\sim 165$
  orbital periods.  The dynamics of models with high opacity is very
  similar to that of purely hydrodynamics models, and it is
  characterized by regular oscillations which are visible also in the
  light curves. As the opacity is decreased, the tori quickly and
  violently migrate towards the gas-pressure dominated regime,
  collapsing towards the equatorial plane. When the spectra of the
  $L_2$ norm of the mass density are considered, high frequency
  inertial-acoustic modes of oscillations are detected (with the
    fundamental mode at a frequency $68 M_{\rm BH}^{-1}\,\rm Hz$), in close
  analogy to the phenomenology of purely hydrodynamic models. An
  additional mode of oscillation, at a frequency $129
  M_{\rm BH}^{-1}\,\rm Hz$, is also found, which can be unambiguously
  attributed to the radiation. The spectra extracted from the light
  curves are typically more noisy, indicating that in a real
  observation such modes would not be easily detected.
\end{abstract}


%====================================================================
\section{Introduction}
\label{s.introduction}
Starting from the seminal papers by  \cite{Shakura1973} and \cite{Novikov:1973},
the physics of accretion discs around black holes has 
been extensively studied, both on a theoretical ground
and by means of numerical simulations 
\citep[see, e.g.,][and references therein]{Abramowicz2011}.
In the last two decades numerical studies related to accretion discs 
have allowed to obtain a good
understanding of complex phenomena such as the development of hydrodynamical and magnetohydrodynamical instabilities, 
the generation and propagation of jets in magnetically dominated regions, the actual role of
the Blandford \& Znajek mechanism, the dynamics of tilted discs, and so on.
In spite of these progresses, most of
the relativistic numerical investigations carried out so far 
have neglected the role of the radiation field. 
The situation is now rapidly changing, and in
the last years quite a few  time-dependent numerical radiative codes have been
developed and applied to 
various physical systems related to accretion
flows in a relativistic
context~\citep{Farris08,Zanotti2011,Fragile2012,
Roedig2012,McKinney2013b,Sadowski2013,Takahashi2013,Sadowski2014}. 

One of the phenomena that has been attracting  a lot of
attention among theoreticians is represented by the periodic response of
geometrically thick discs to perturbations.  These objects
were introduced almost 40 years ago as stationary and
axisymmetric solutions of a fluid in non-Keplerian
circular motion around a black
hole~\citep{Fishbone76,Abramowicz78,Kozlowski1978}.
Although they can be regarded as toy models under many respects,
they are still considered as a realistic approximation of the inner part of a much 
larger accretion disc.
Their astrophysical relevance was further revitalized  once it
was found that they can exhibit a regular oscillating
behaviour when subject to generic external
perturbations~\citep{Zanotti03, Zanotti05, Blaes2006}. In
fact, the dynamics of geometrically thick discs was thereafter
studied in connection  to the phenomenology of 
the high frequency quasi-periodic oscillations (kHz-QPOs) observed in the
X-ray spectra of binary systems~\citep{Remillard2006}, for which they could
provide more than one plausible interpretation [see,
among the others,
\cite{Rezzolla_qpo_03b,Abramowicz2003,Schnittman06,Blaes2007,Montero2012,Mazur2013,Vincent2014,Bakala2015}].
In this context, \cite{Mishra2015} have recently studied the response 
of a hydrodynamical torus to various perturbations
 and used sophisticated ray-tracing to
obtain luminosity curves assuming both large and small optical depth. 

%ooooooooooooooooooooooooooooooooooooooooooooooooooooooooooooo
\begin{table*}% [!b] 
\begin{center} 
\begin{tabular}{c|ccccccc}
\hline
Name of the Model & $r_{\rm in}$ & $r_{\rm out}$ & $t_{\rm orb}$ & Entropy & $T_{\rm c}$ & $\rho_{\rm c}$ & $\tau_{\rm sc}$ \\
\hline
  & [M] & [M] & [${\rm s}$] & [{\rm geo}] & [${\rm K}$] & [${\rm g}\,{\rm cm}^{-3}$] & \\
\hline
Hydro43 & 6.54 & 11.00 & $7.5\times10^{-3}$  & 100 & 4.82e9 & 1.63e8 & n/a \\
Hydro53 & -    & -     & -                 & 100 & 9.65e9 & 1.59e8 &  n/a \\
Rad10   & -    & -     & -                 & 10  & 6.61e8 & 1.07e2 & 4.68e8 \\
Rad30   & -    & -     & -                 & 30  & 2.95e8 & 3.88e1 & 1.73e7 \\
Rad100  & -    & -     & -                 & 100 & 4.26e7 & 1.05e0 & 4.68e5 \\
Rad400  & -    & -     & -                 & 400 & 4.23e7 & 1.64e-2 & 7.31e3 \\
\hline
\end{tabular}
\caption{Initial conditions of the considered tori. The
  mass of the black hole is assumed to be
  $M_{\rm BH}=10M_{\odot}$. The columns report: the inner and
  outer radii of the torus, the orbital period at the
  maximum density radius $r_c=8.35$, the specific entropy (in $c=G=1$ units), the
  temperature, the rest-mass density and the scattering
  opacity, the last three computed at $r_c$. }
\end{center}
\label{tab:tori}
\end{table*}

In this paper we wish to perform a step forward along this
  direction of research by studying the role of the radiation field on
  the very oscillations of radiation pressure-dominated tori.  To
this extent, we resort to the radiative general relativistic code
\texttt{KORAL}~\citep{Sadowski2013,Sadowski2014}, which incorporates a
number of very desirable features, such as i) the conservative
formulation of the equations, ii) the implicit treatment of radiative
source terms and iii) the ability in capturing both the optically
thick and the optically thin regimes, by means of a suitable
relativistic version of the M1-closure scheme.  After considering the
long-term evolution of a few but representative models in two spatial
dimensions, we find that the dynamics of the tori is strongly
dependent on the opacity and produces light curves which contain the
in-printing of the perturbation.  Although in our analysis we do not
take into account important radiation processes like bremsstrahlung or
synchrotron, our results testify the importance of a proper physical
and numerical modelling of radiation processes in a relativistic
context. We emphasize that we do not aim at proposing any specific new
  model for the kHz-QPOs phenomenology, but rather we extend the
  numerical study of oscillating tori to optically thick,
  radiation-pressure supported objects.

In the following, we adopt a geometrized system of units by setting $c=G=1$, although we resort to ${\rm cgs}$ units 
when convenient. In particular, we adopt $r_{\rm
  g}=GM/c^2$ and $t_{\rm
  g}=GM/c^3$ as units of distance and time, respectively.

%==================================================================== 
\section{Numerical setup}
\label{s.numerical}
\subsection{Initial configuration}
Constant angular momentum tori, which are stationary and axisymmetric solutions of the relativistic hydrodynamics equations
for a perfect fluid can be built following \cite{Kozlowski1978} [see also \cite{Font02a}]. 
The resulting hydrodynamic model provides the distribution of the total
pressure, $p_{\rm tot}$, of the rest-mass density $\rho$, as well as of the orbital velocity $\Omega$, which is 
super-Keplerian in the inner part of the torus up to the radius of maximum density, while it is sub-Keplerian in the outer part.
The gas obeys the ideal gas equation of state
$p=\rho\epsilon(\Gamma-1)$, where $\Gamma$ is the
adiabatic index and $\epsilon$ is the specific (i.e. per
unit mass) internal energy.

We decided to simulate the oscillations of a class of relativistic tori
described by constant specific angular momentum $\ell=-u_\phi/u_t=3.8$
with the inner and outer edges at $r_{\rm in}=6.54$ and $r_{\rm
  out}=11.0$, respectively.  The jump of the effective potential $W$
between $r_{\rm in}$ and the position of the cusp is
$W_{{\rm in}}-W_{{\rm cusp}}=-7.37\times10^{-3}$,
meaning that the equilibrium configuration remains below the
marginally stable one [see \cite{Font02a} for more details].  The
density of the torus is adjusted through the entropy parameter, which also affects the Thomson scattering opacity.
Specific values are given in Table~\ref{tab:tori}.

When radiation is also considered, we assume that initially $\Gamma=4/3$, consistently with the assumption of a radiation pressure
dominated disc. The total pressure obtained in this way is then distributed
between the gas and the radiation, in such a way to satisfy  local thermal equilibrium
(LTE), i.e. $\widehat E=4\sigma T^4$. This involves solving the following quartic equation
for gas (and radiation) temperature \citep{Sadowski2014}
\be 
p_{\rm tot}=p_{\rm gas}+p_{\rm rad}=k_{\rm B}\rho T + \frac 43 \sigma T^4\,,
\ee 
where $k_{\rm B}$ and $\sigma$ are the Boltzmann and the Stefan-Boltzmann constants, respectively.
Once the initial configuration has been computed, the polytropic index of the non-relativistic 
gas is fixed at $\Gamma=5/3$ during the simulations. 
We emphasize that 
the tori are initially not in perfect equilibrium.\footnote{A truly equilibrium model for a thermodynamic torus around a black hole, which is however 
not valid for ideal gas equations of state, has been described in \cite{Zanotti2014}.}
This is due to the fact that the effective polytropic index of the gas and radiation
mixture may not be exactly $4/3$, and because in constructing the
torus radiation field we have neglected the radiation pressure force
along the radiative flux
However, as long as we are interested in the time-varying behaviour of the tori, this should not be regarded as a weak point.  
Actually, in order to trigger the development of oscillations, all the models 
are perturbed by addition of a small radial velocity component, which is a rescaling of the spherical accretion solution by \cite{michel72}, i.e.
\begin{equation}
v^r = \eta \,v^r_{\rm Michel}\,.
\end{equation}
In practice, we act like in \cite{Zanotti03}, choosing a perturbation amplitude $\eta=0.03$. A detailed study about the effects of different directions of the initial velocity perturbation has been presented by \cite{Mishra2015} for hydrodynamic slender tori.

%\\[2ex]

%
\subsection{Numerical approach}
%

%%%%%%%%%%%%%%%%%%%%%%%%%%%%%%%%%%%%%%%%%%%%%%%%%%%%%%%%%%%%%%%%%%%%%%%%%%%%%%%%%%%%%%%%%

\begin{figure*}
	\centering
	\begin{tabular}{cc}
		%Entropy30
		\includegraphics[width=0.48\textwidth,height=0.19\textheight]{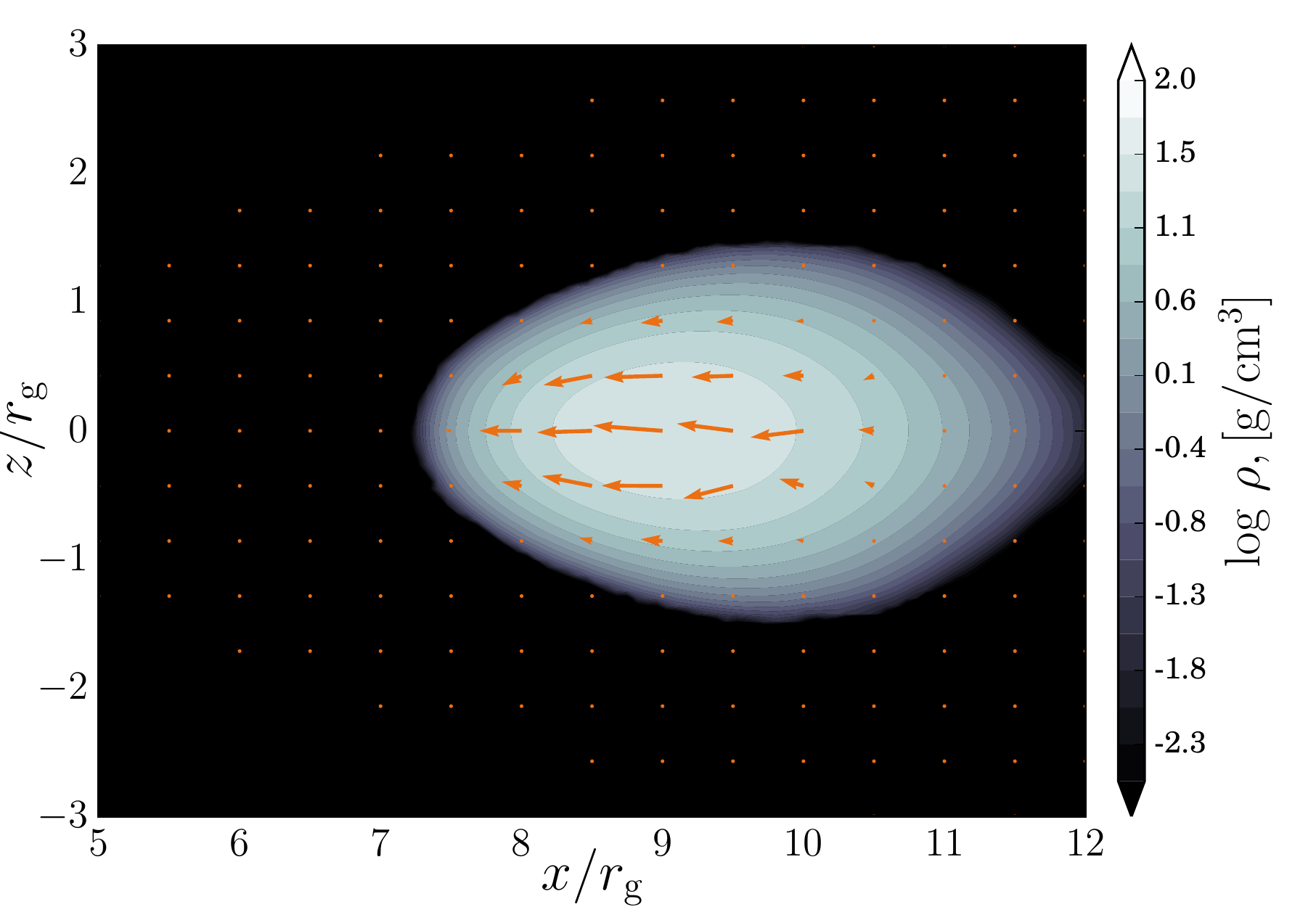} &
		\includegraphics[width=0.48\textwidth,height=0.19\textheight]{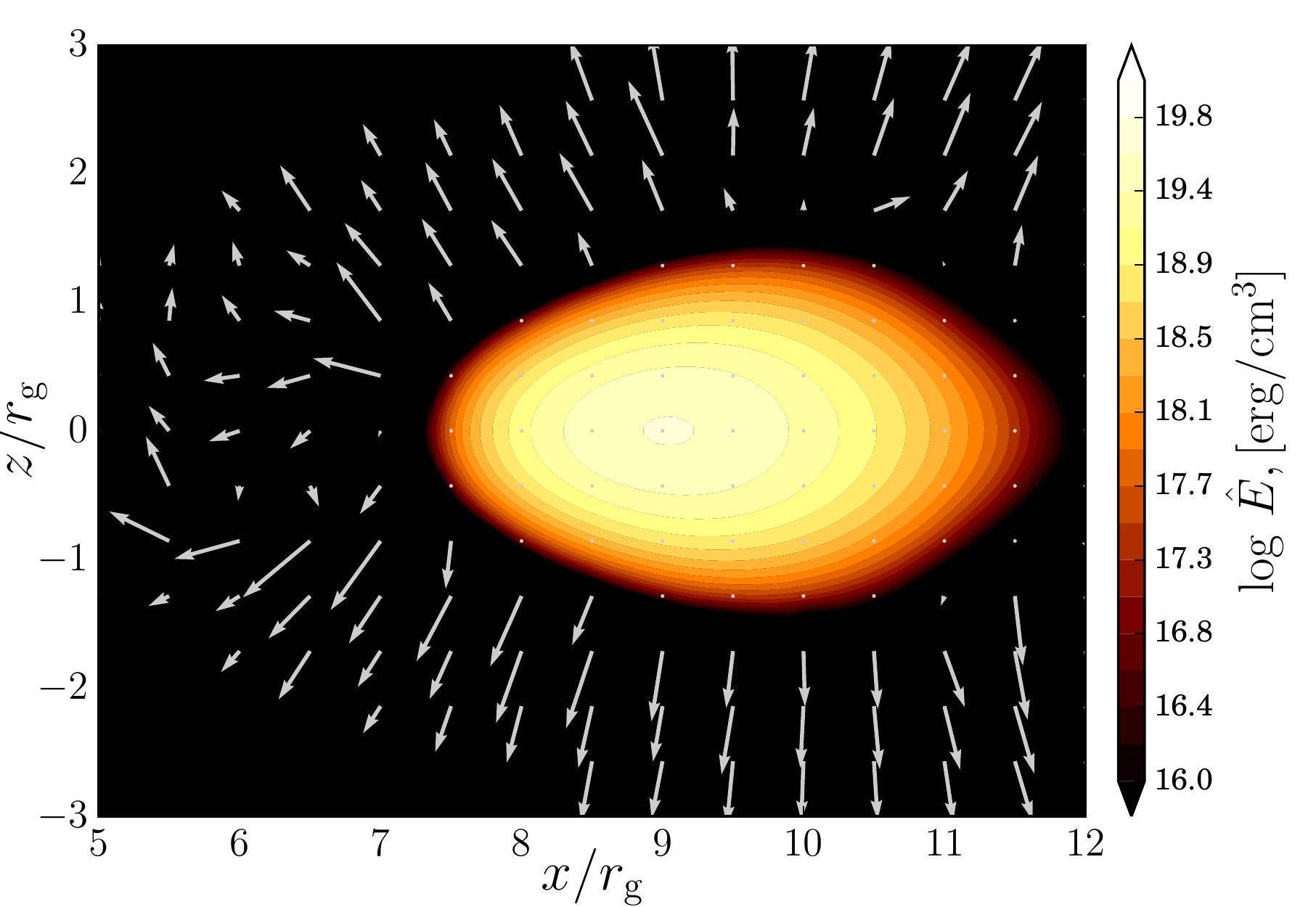} \\
		\includegraphics[width=0.48\textwidth,height=0.19\textheight]{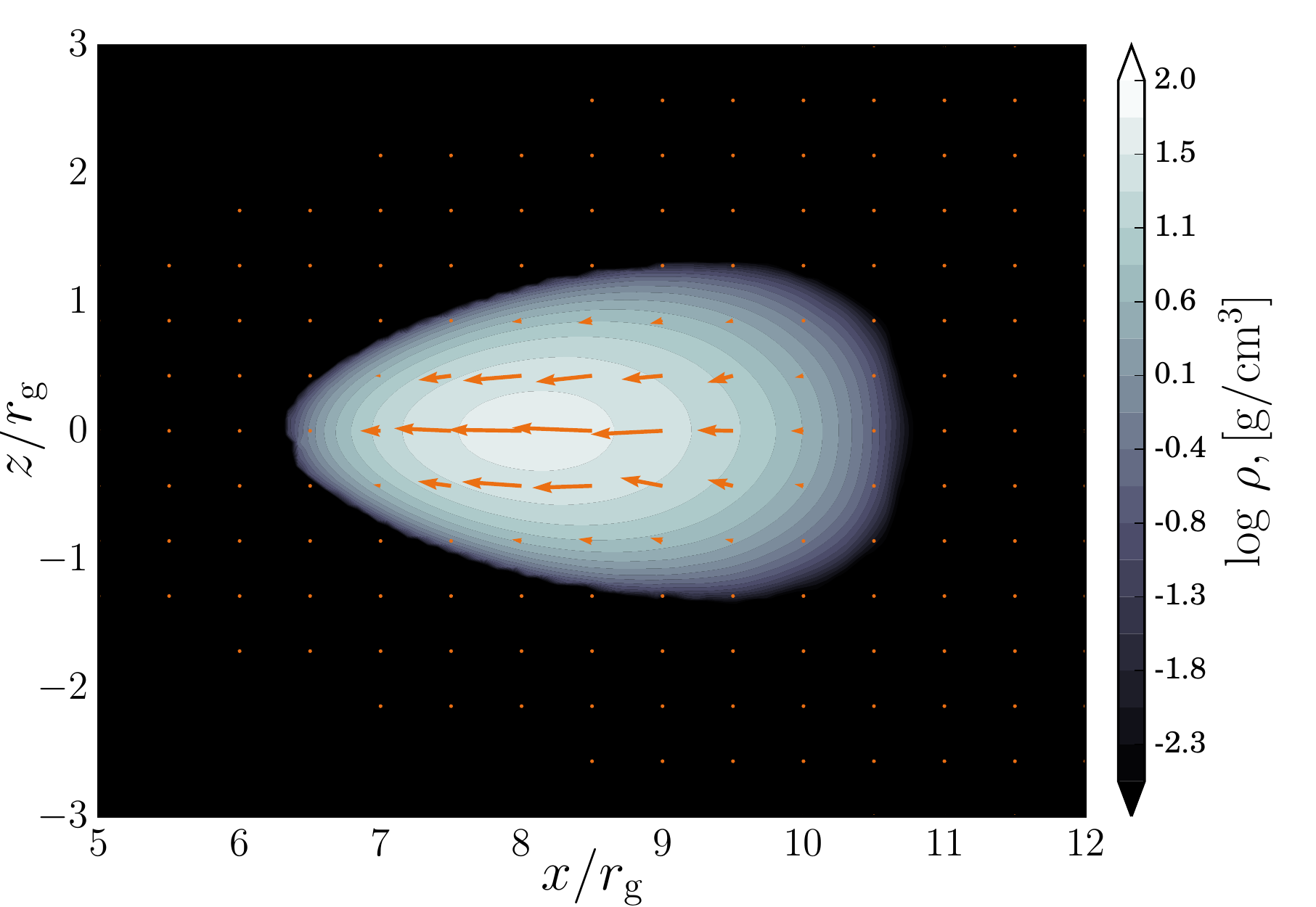} &
		\includegraphics[width=0.48\textwidth,height=0.19\textheight]{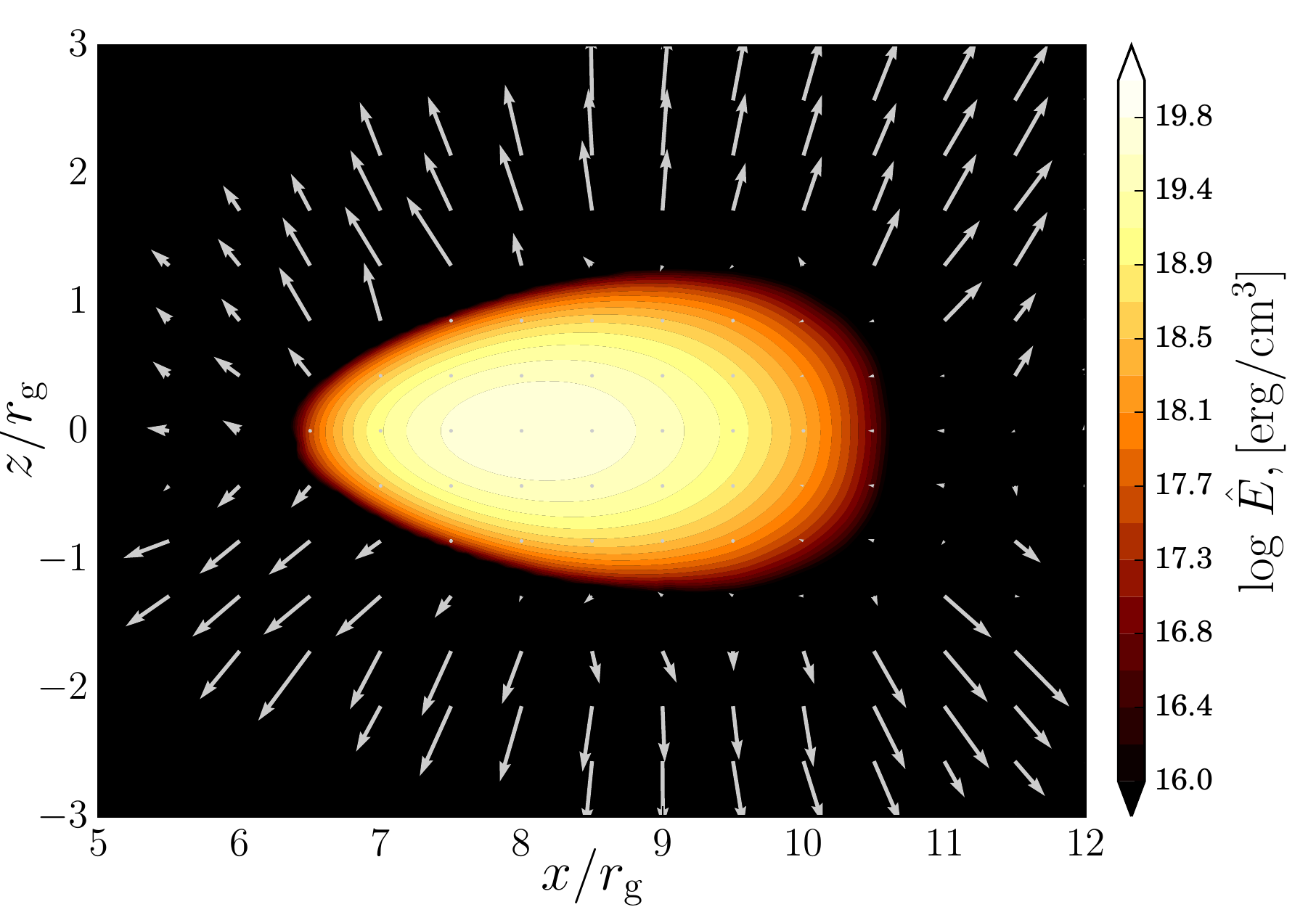} \\
		\includegraphics[width=0.48\textwidth,height=0.19\textheight]{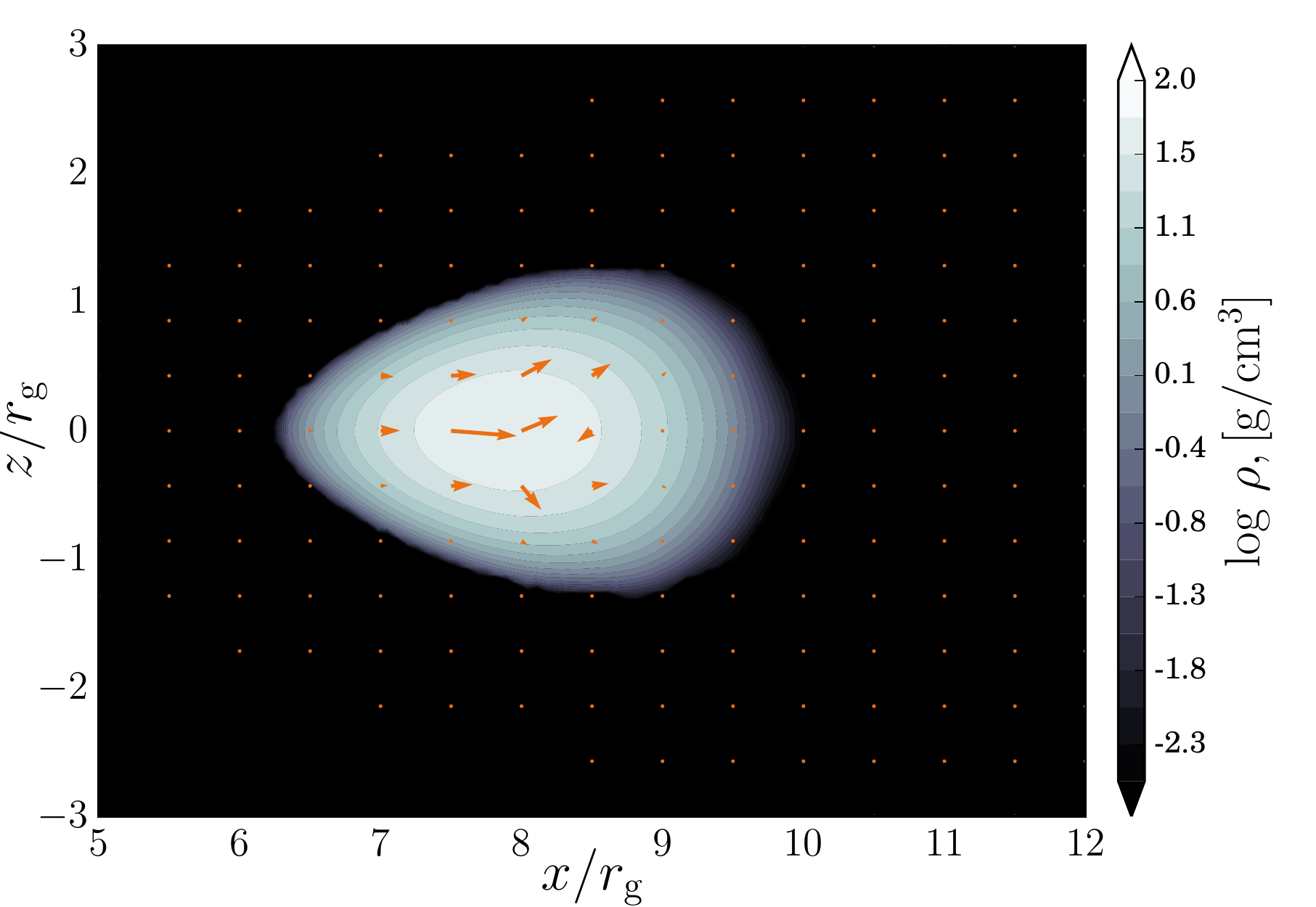} &
		\includegraphics[width=0.48\textwidth,height=0.19\textheight]{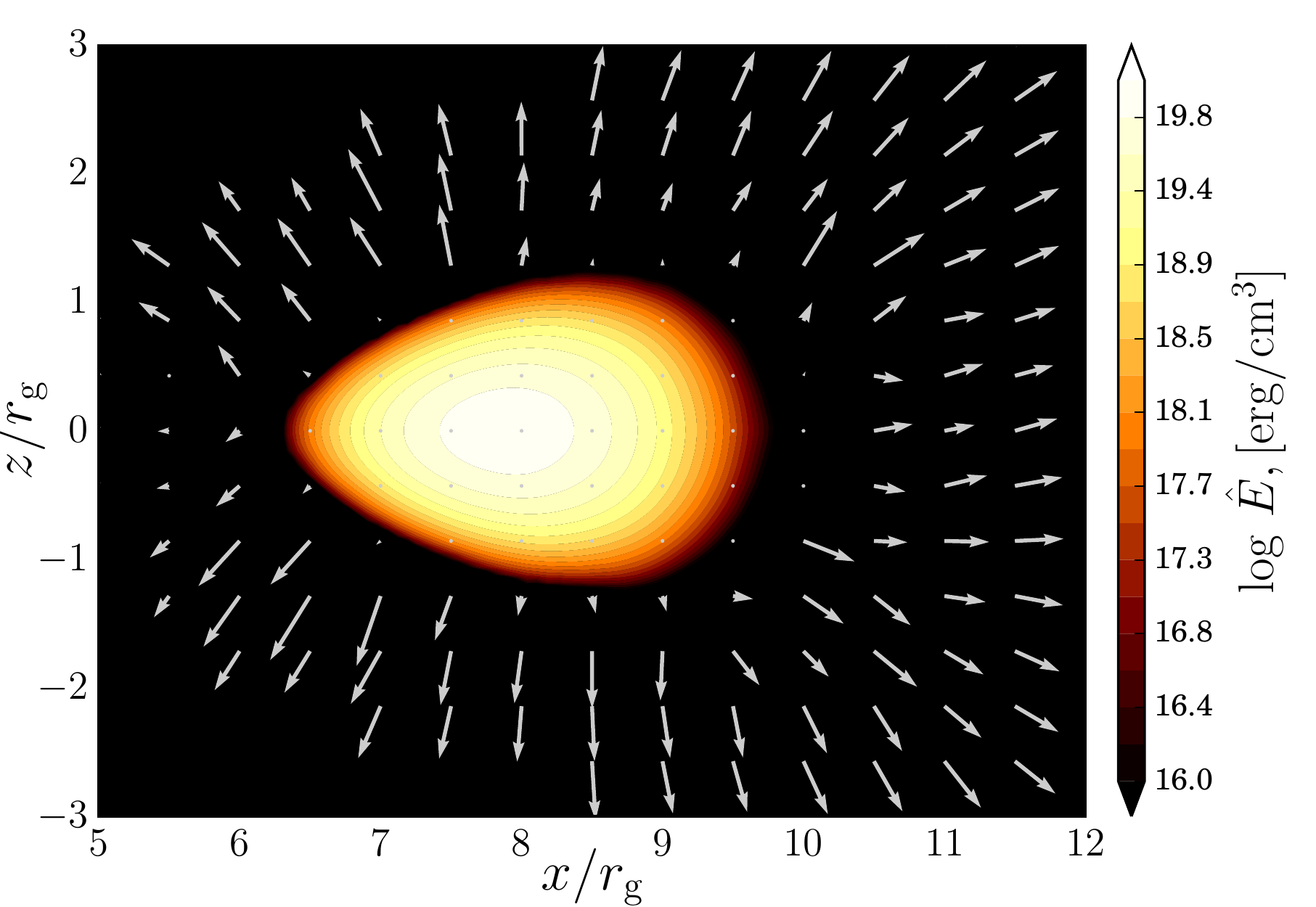} \\
		\includegraphics[width=0.48\textwidth,height=0.19\textheight]{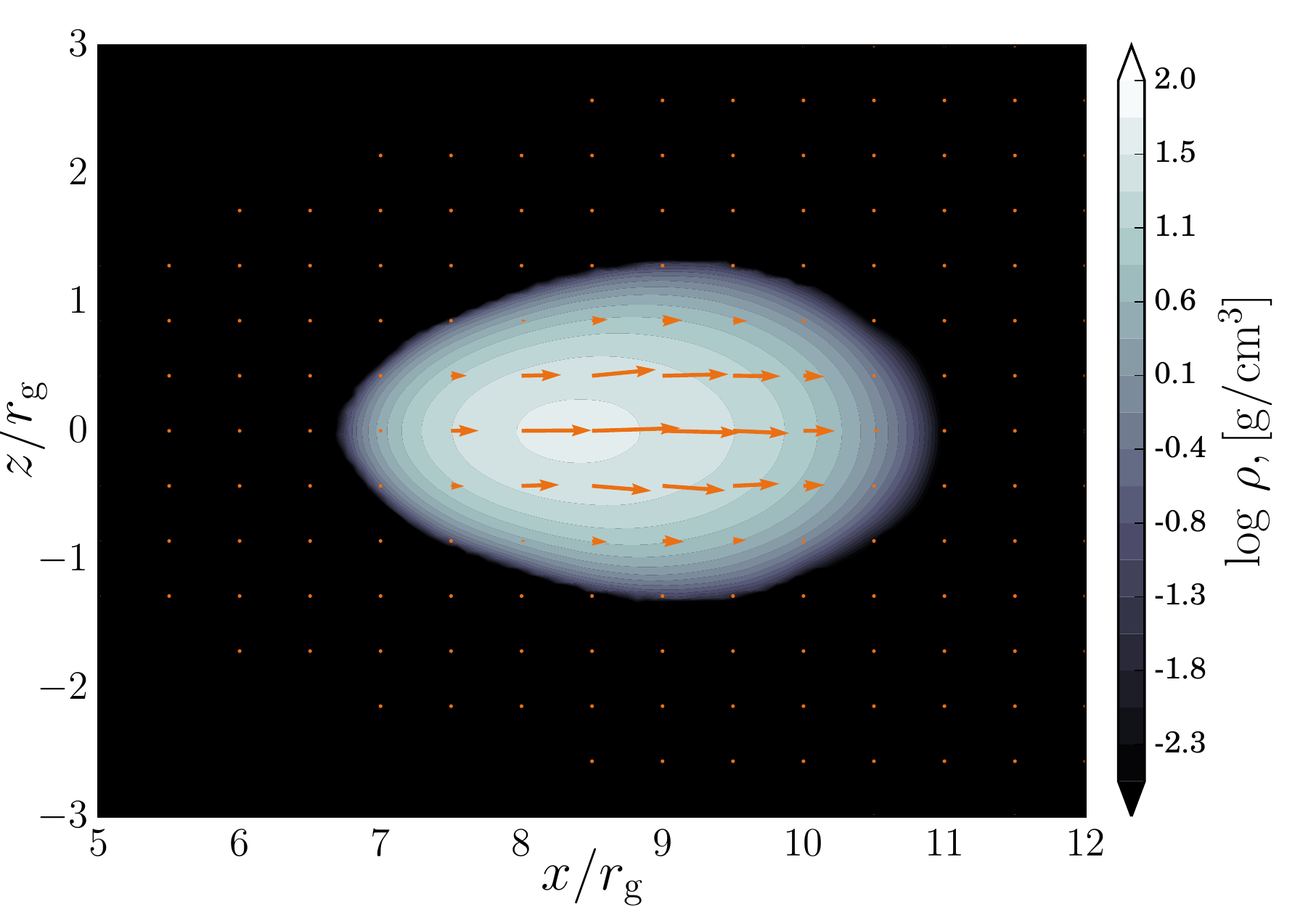} &
		\includegraphics[width=0.48\textwidth,height=0.19\textheight]{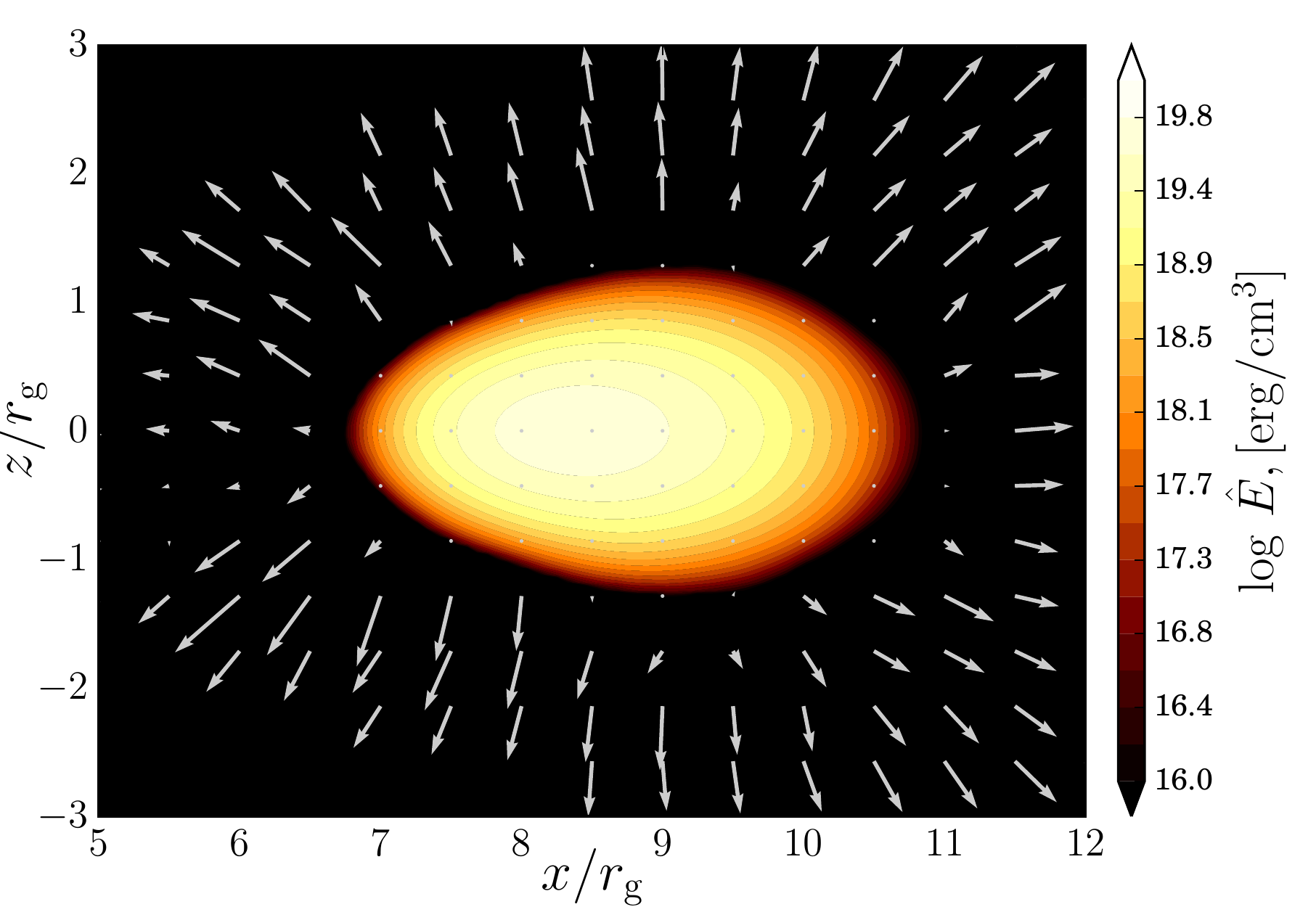} \\
		\includegraphics[width=0.48\textwidth,height=0.19\textheight]{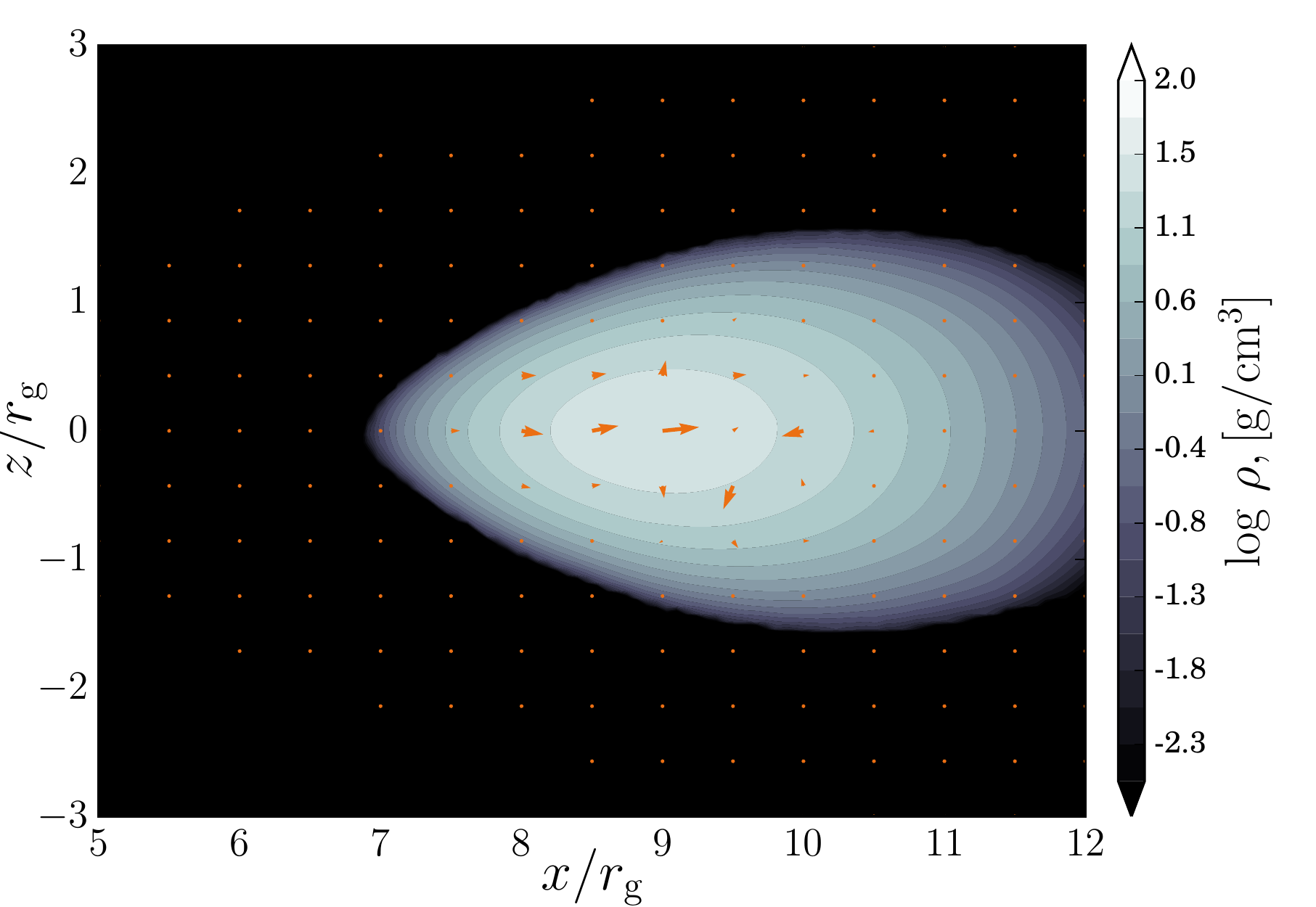} &
		\includegraphics[width=0.48\textwidth,height=0.19\textheight]{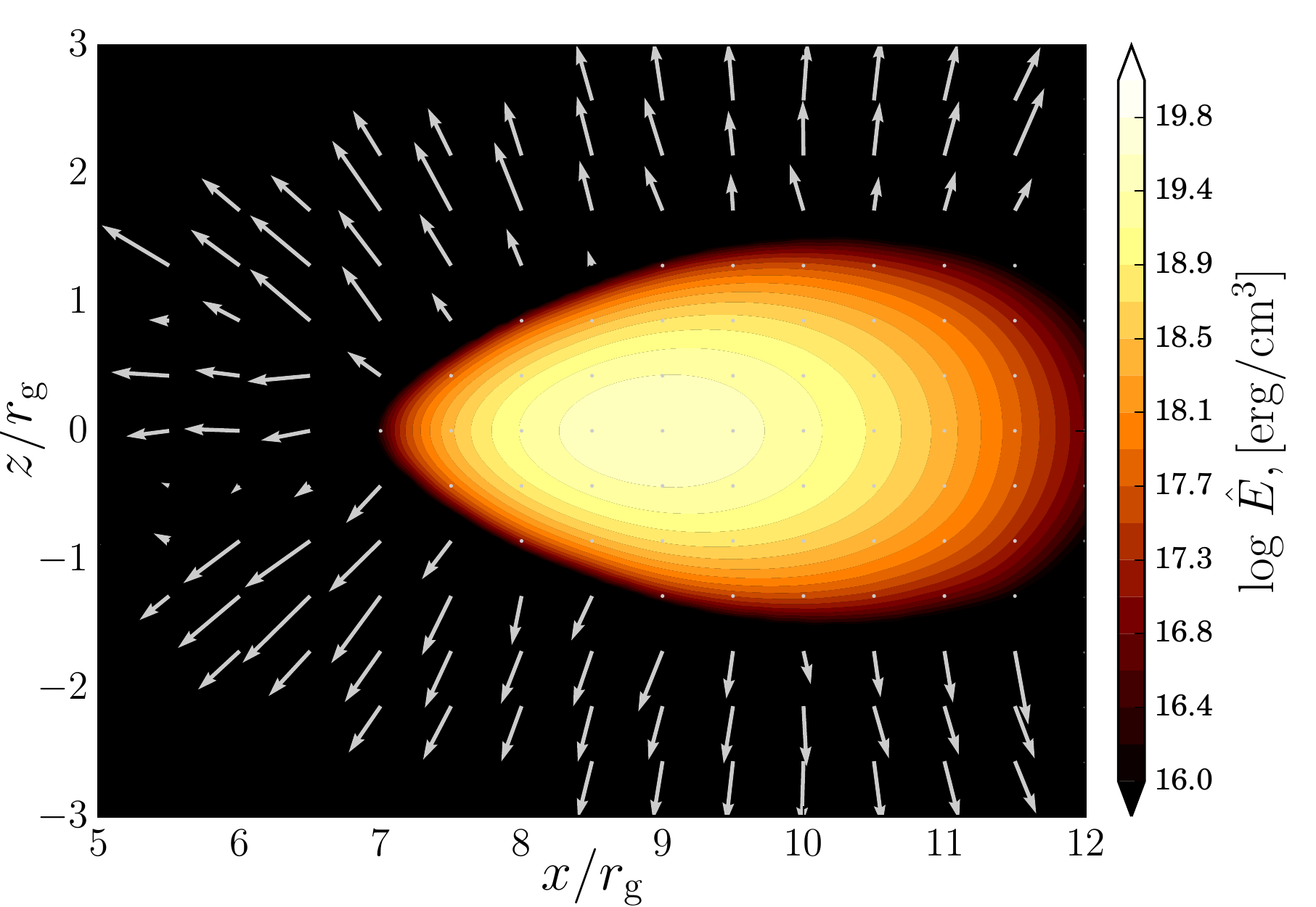} \\
	\end{tabular}
	\caption{Illustration of a typical oscillation sequence for the model Rad30 through six snapshots (from top to bottom). Left panels: distribution of the rest-mass density 
	with the velocity represented by orange arrows.
	Right panels: distribution of the energy density of the radiation field, with the energy flux represented by  white arrows. For illustrative purposes, the $x-$ axis has been cut at 12.}
	\label{fig:osequence}		
\end{figure*}
%%%%%%%%%%%%%%%%%%%%%%%%%%%%%%%%%%%%%%%%%%%%%%%%%%%%%%%%%%%%%%%%%%%%%%%%%%%%%%%%%%%%%%%%%%

%ooooooooooooooooooooooooooooooooooooooooooooooooooooooooooooo
\begin{figure*}
	\centering
	\begin{tabular}{ccc}
		%Hydro53
		%\includegraphics[width=0.32\textwidth]{Hydro530001.pdf} &
		\includegraphics[width=0.32\textwidth]{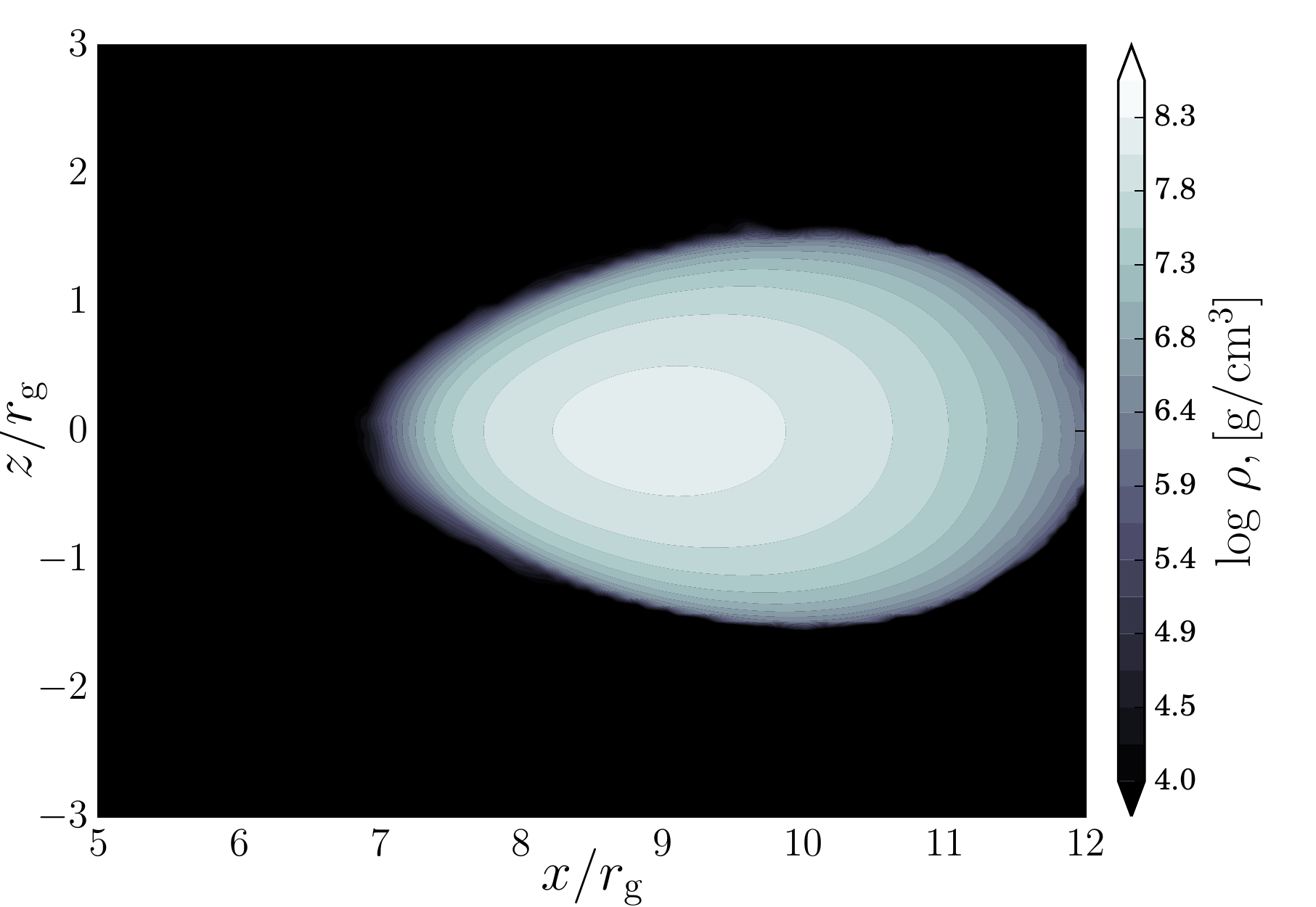} &
		\includegraphics[width=0.32\textwidth]{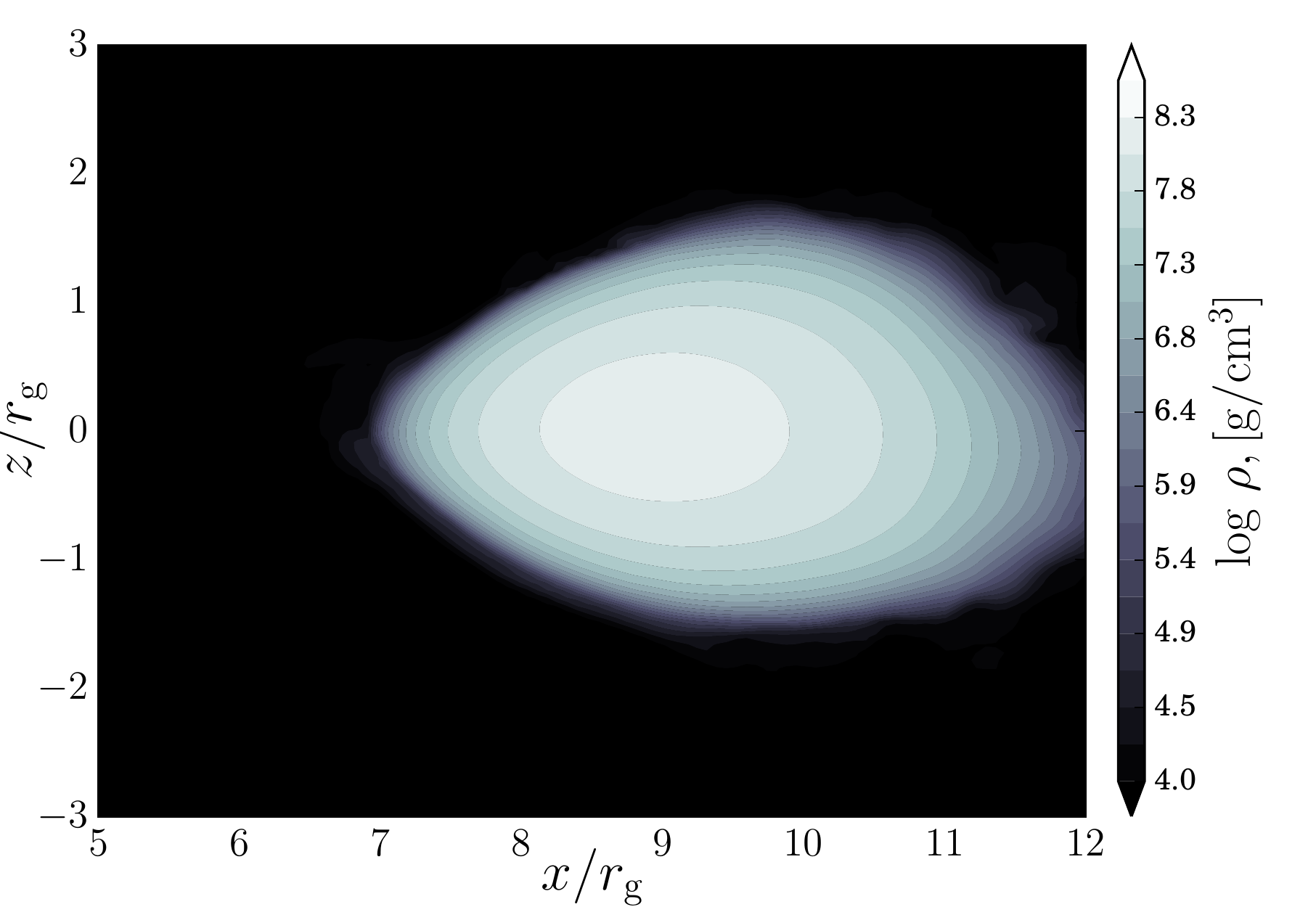} &
		\includegraphics[width=0.32\textwidth]{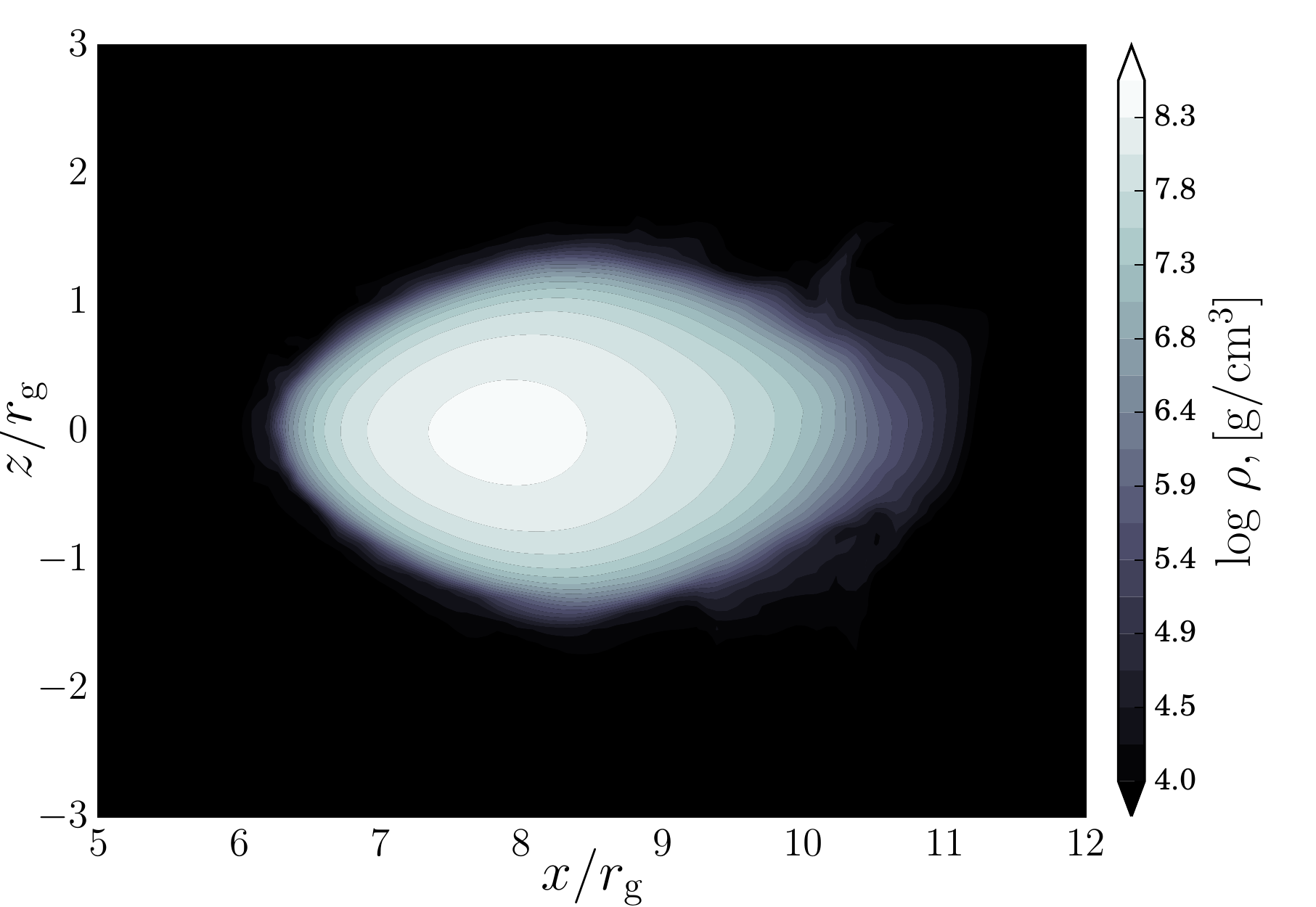} \\
		%\vspace{0.9cm}
		%Entropy30
		%\includegraphics[width=0.32\textwidth]{Rad3000001.pdf} &
		\includegraphics[width=0.32\textwidth]{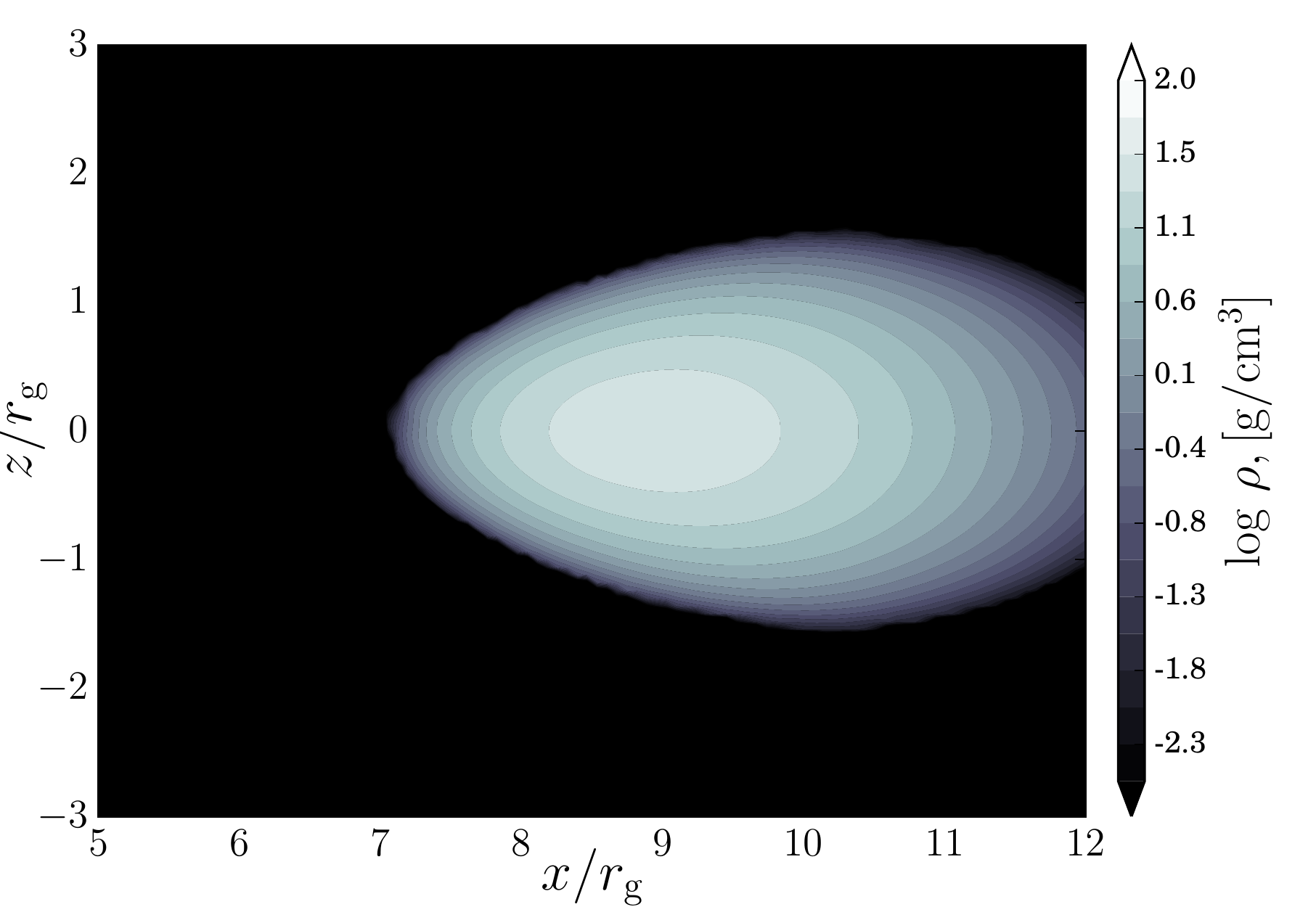} &
		\includegraphics[width=0.32\textwidth]{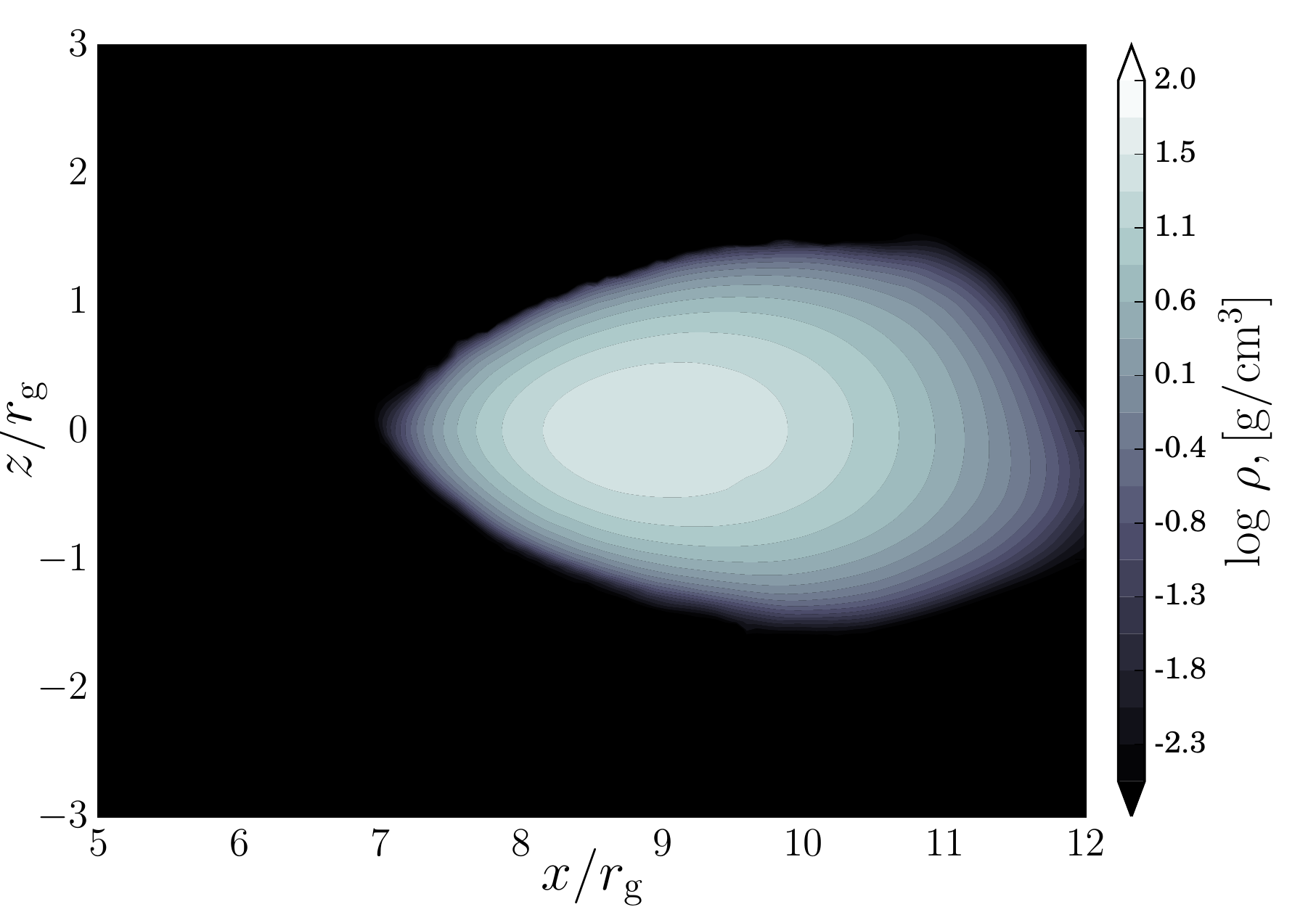} &
		\includegraphics[width=0.32\textwidth]{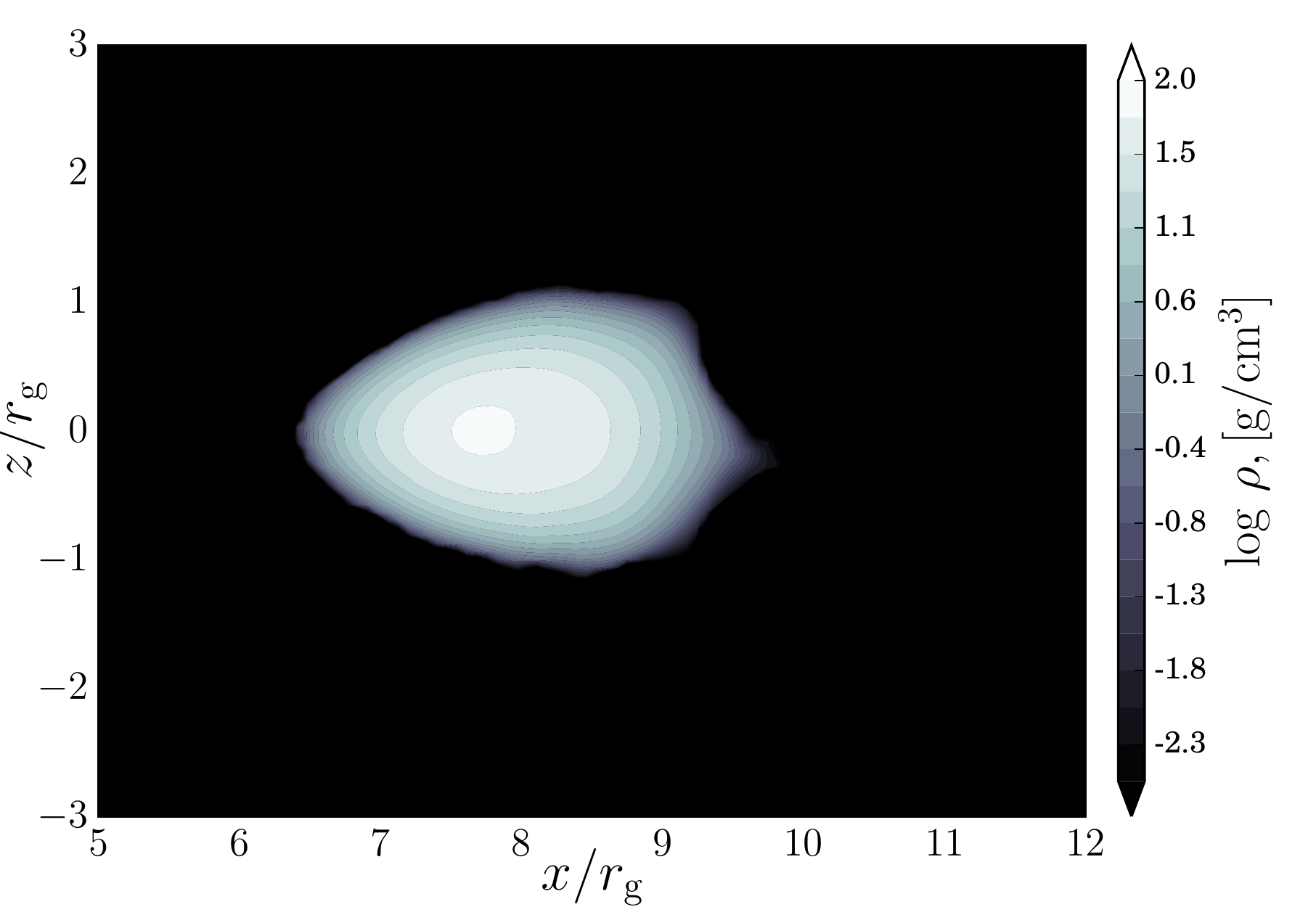} \\
		%Entropy400
		%\includegraphics[width=0.32\textwidth]{Rad4000001.pdf} &
		\includegraphics[width=0.32\textwidth]{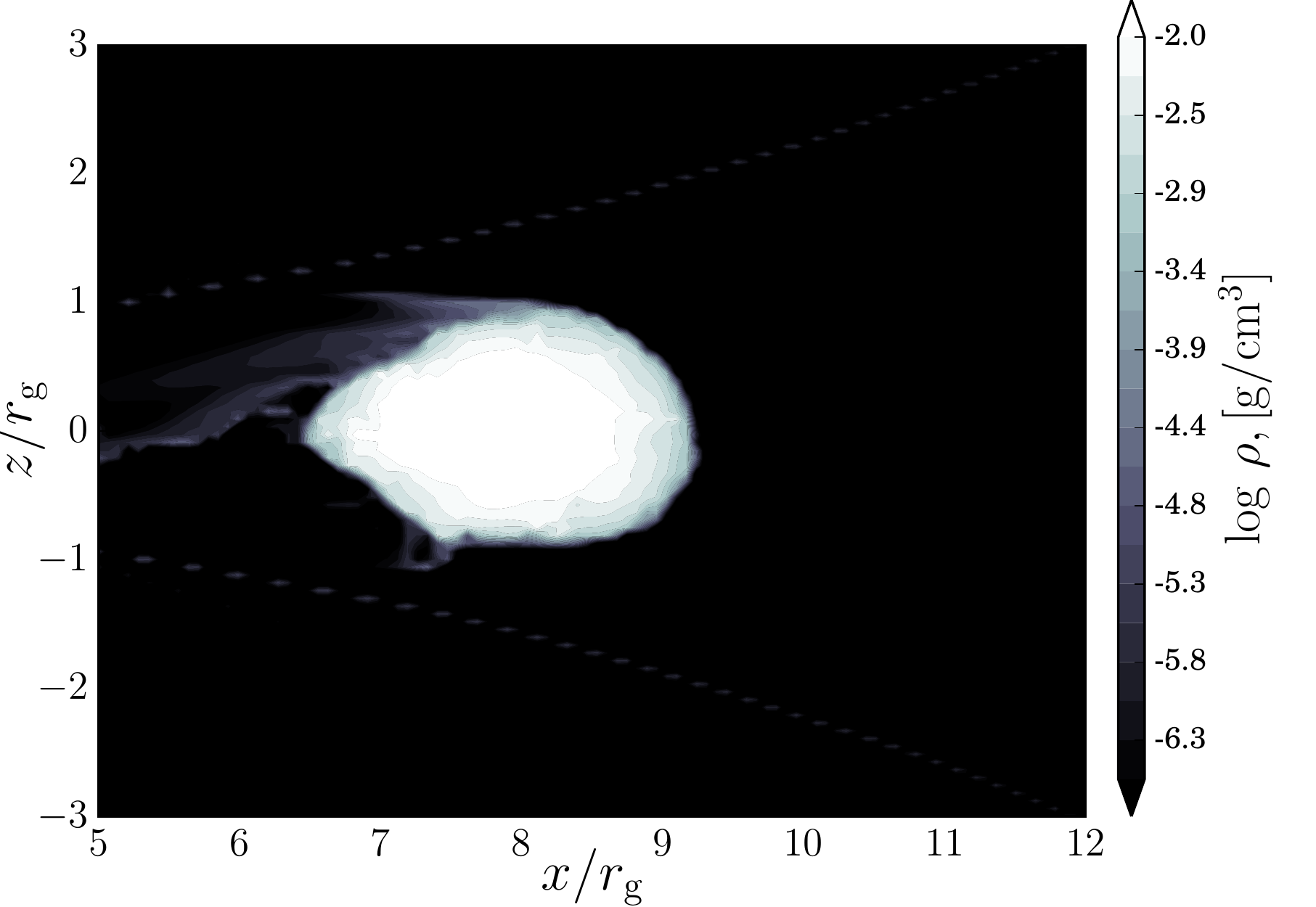} &
		\includegraphics[width=0.32\textwidth]{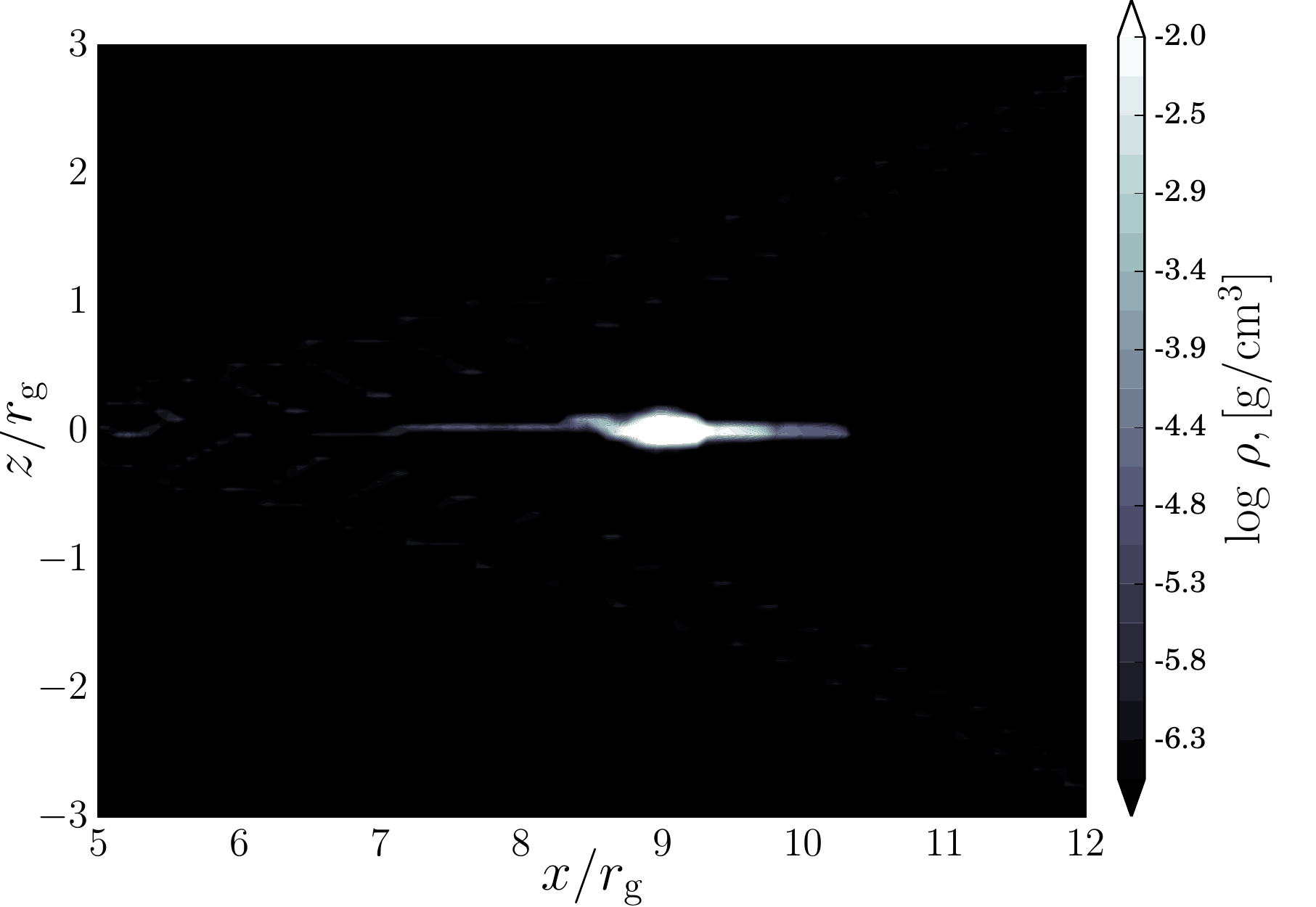} &
		\includegraphics[width=0.32\textwidth]{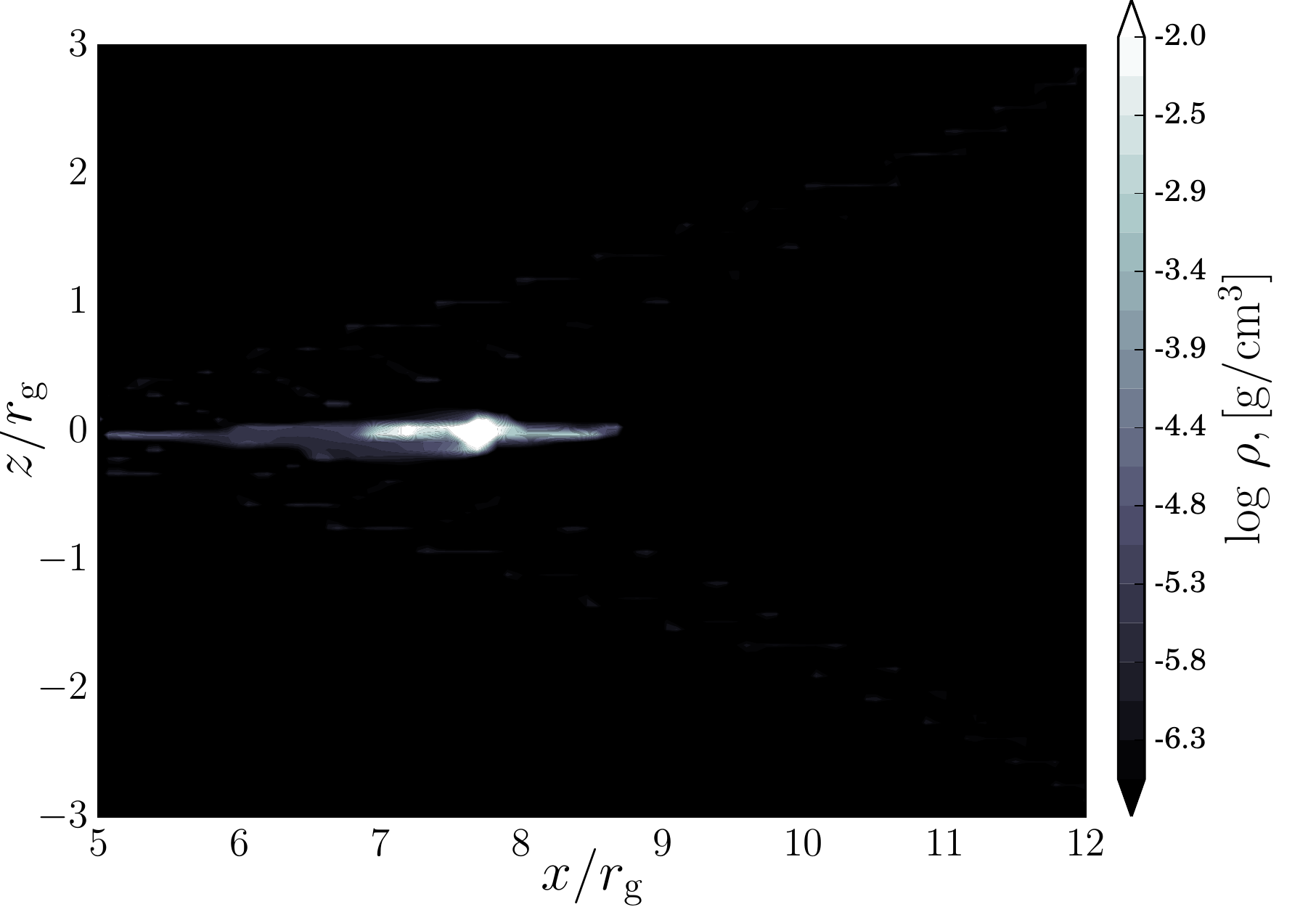} \\
	\end{tabular}
	\caption{Rest-mass density distribution of the models Hydro53, Rad30 and Rad400, from top to bottom. First column: 
snapshot at $t\approx 1000\approx 0.049$\,s. Second column: snapshot at $t\approx 10000\approx 0.49$\,s. Last column: snapshot at the end of the simulation. The snapshots are chosen in such a way to represent the same phase of oscillation for each model. The density profile range covers 4.5 orders of magnitude. The density scale for the model Hydro53 is different from the other two.}
	%First row: Hydro53. Snapshots at: 0001,1999,4953. Second row: Rad30.Snapshots at 0001, 1998, 4951. Third Row: Rad400 Snapshots at 0001 1986. Columns set in order 0001, 2001, 5000}
	\label{fig:3models}		
\end{figure*}
%%%%%%%%%%%%%%%%%%%%%%%%%%%%%%%%%%%%%%%%%%%%%%%%%%%%%%%%%%%%%%%\newpage

We solve the general relativistic radiation-hydrodynamics equations in a Schwarzschild
metric using the \texttt{KORAL} code
~\citep{Sadowski2013,Sadowski2014} which evolves the gas
component,
the radiation field, and the interaction between them in a conservative
way using Godunov, finite-difference methods. The radiation is
evolved adopting the M1 closure scheme \citep{levermore+84}. 

The simulations assume axisymmetry and run in $2.5$ dimensions, that
is, allowing for non-zero azimuthal vector
components.  We assume a black hole mass of $10 M_{\odot}$ and evolve the models for
roughly $1.2s$, corresponding to $\sim 165$ orbital periods at the
pressure maximum of the initial model. The numerical domain extends from $r_{\rm min}=5$ to $r_{\rm max}=14$ along the radial direction, and 
from $\theta=\pi/2-0.3$ to $\theta=\pi/2+0.3$ along the $\theta$ direction, and it is covered by a
uniform (in radius and polar angle) grid composed of 
$448\times448$ grid-points. We note that the radial resolution is therefore $\Delta r=0.02$.
Standard outflow/no-inflow boundary conditions are adopted at all the
boundaries\footnote{Such a setup effectively prevents the torus from
  being self-illuminated, which could in principle affect (and
  complicate) its dynamical behaviour.}.

For simplicity, in this work we adopt  only the scattering opacity
$\kappa_s = 0.34 ~\mathrm{cm^2/g}$, and we account for the
Comptonization using the ``black-body'' method
\citep[see][]{sadowski+compt}. Since we do not assume any external heating source,
the tori are expected to cool down by emitting the radiation that supports them.

\section{Results}
%
%================================================================
\subsection{Torus dynamics}
All the models have been evolved up to $t=25000$, corresponding to $\sim 165$ orbital periods computed at the 
maximum density radius $r_c=8.35$.
Two purely hydrodynamic models, namely Hydro43 and Hydro53, have been also studied, in order to provide a comparison with respect to the radiative models, which is especially useful in the spectral analysis of Sect.~\ref{sec:spectral}.

Because of the radial perturbation introduced, the tori start oscillating in the potential well. However, since the initial models are significantly below the
marginally stable configuration, no appreciable mass accretion is induced by the oscillations and the periodic behaviour can last for a long time.
In Fig.~\ref{fig:osequence} a full oscillation sequence at $t\sim
0.1$\,s is monitored through six snapshots  for the model Rad30. The
left and the right panels report the distribution of the rest-mass
density and of the radiation energy density, respectively. The arrows
in the radiative panels reflect the free-streaming radiation leaving
the torus surface. As expected, during the 
evolution the energy density of the radiation remains concentrated
towards the center of the torus, where the optical depth is the
largest. Radiation diffuses out following the gradient of radiative
energy density and ultimately leaves the surface of the torus. The
distribution of the out coming radiation is determined by the shape of
the photosphere. In the case of the tori studied here, most of
the radiation is emitted roughly in the direction perpendicular to the
equatorial plane. 

\begin{figure}
	\centering
	\includegraphics[width=1.075\columnwidth]{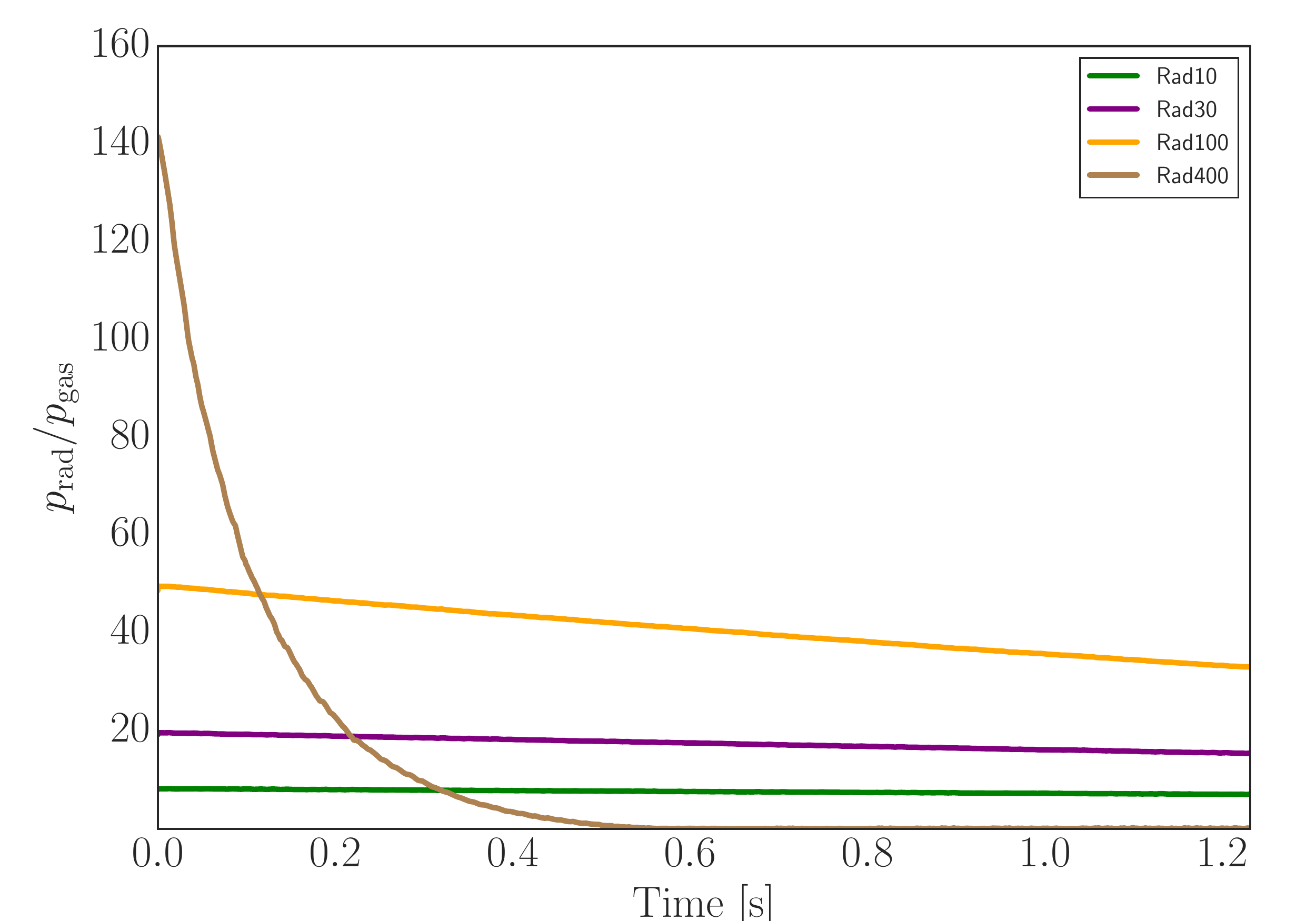}
	\caption{Radiation-to-gas pressure ratio as a function of time. Systems with the lowest entropy are preserving the ratio $p_{\rm rad}/p_{\rm gas}$ for the whole evolution. A drastic transition  is observed for the model Rad400.}
	\label{fig:pratio}
	\end{figure}

Additional insight on the dynamics can be inferred from Fig.~\ref{fig:3models}, whose rows compare the hydrodynamic model Hydro53 (top row), with the two radiation models Rad30 
(middle row) and Rad400 (bottom row). We recall that at time $t=0$ the three models have the same size. 
The most dramatic behaviour is exhibited by the highest entropy and
lowest optical depth  model
Rad400, which can be regarded as an extreme case.
Initially, this model is the  most radiation-pressure dominated  but 
after 80 orbital periods it has lost most of its radiation, entering a phase of pure hydrodynamic evolution. 
Fig.~\ref{fig:3models} can indeed be interpreted in parallel with the information provided by Fig~\ref{fig:pratio}, where we have reported, for each of the radiative models, the evolution of the radiation-to-pressure ratio. The lower the specific entropy (see Table~\ref{tab:tori}), the smaller is the initial value
of $p_{\rm rad}/p_{\rm gas}$, but, simultaneously, the slower is the decay of $p_{\rm rad}/p_{\rm gas}$ during the evolution, since more photons are trapped in the more optically thick gas. As a result, while the models Rad10 and Rad30 preserve most of their radiation content for tens of orbital periods, the model Rad400, with a smaller optical thickness, quickly evaporates by radiation emission, collapsing towards a much smaller structure along the equatorial plane, as evident from the last row of Fig.~\ref{fig:3models}.

\begin{figure}
	\centering
	\includegraphics[width=1.075\columnwidth]{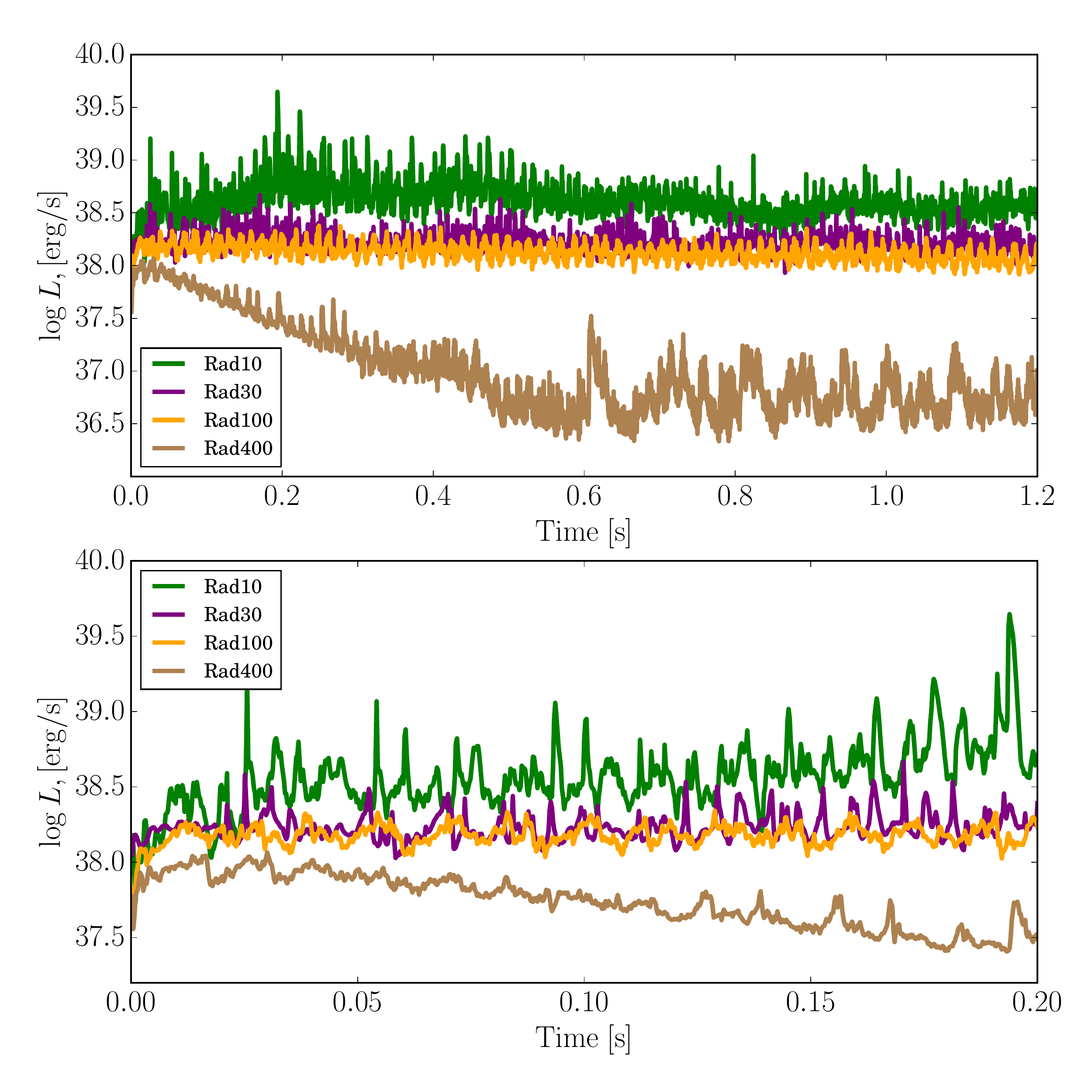}
	\caption{Top panel: Luminosity as a function of time for the full simulation period. Bottom panel: Zoom of the luminosity during the first 0.2\,s.}
	\label{fig:lum}
\end{figure}

The energy outcome from the oscillating tori is illustrated in
Fig.~\ref{fig:lum}, where we have reported the luminosity, in units of
${\rm erg\,s^{-1}}$, and computed by means of a simple integral of the
radiative fluxes through a suitable outer surface that encloses the
whole oscillating torus. 
This quantity is a measure of
  radiation emitted by the torus, but it does not exactly correspond to
the luminosity reaching an observer, for which the ray-tracing of photons would be needed \citep[see][for such an approach in the purely hydrodynamic framework]{Mishra2015}. 
There is a clear periodic pattern in the
light curves that we obtain, which we will analyze more closely in the next Section.
While collapsing towards the equatorial plane, 
  the model Rad400 cools down quickly and loses the radiation that supported it, until it reaches a new
  quasi-stationary regime characterized by a luminosity one order of
  magnitude smaller than the initial value.

\begin{figure}
	\centering
	\includegraphics[width=1.075\columnwidth]{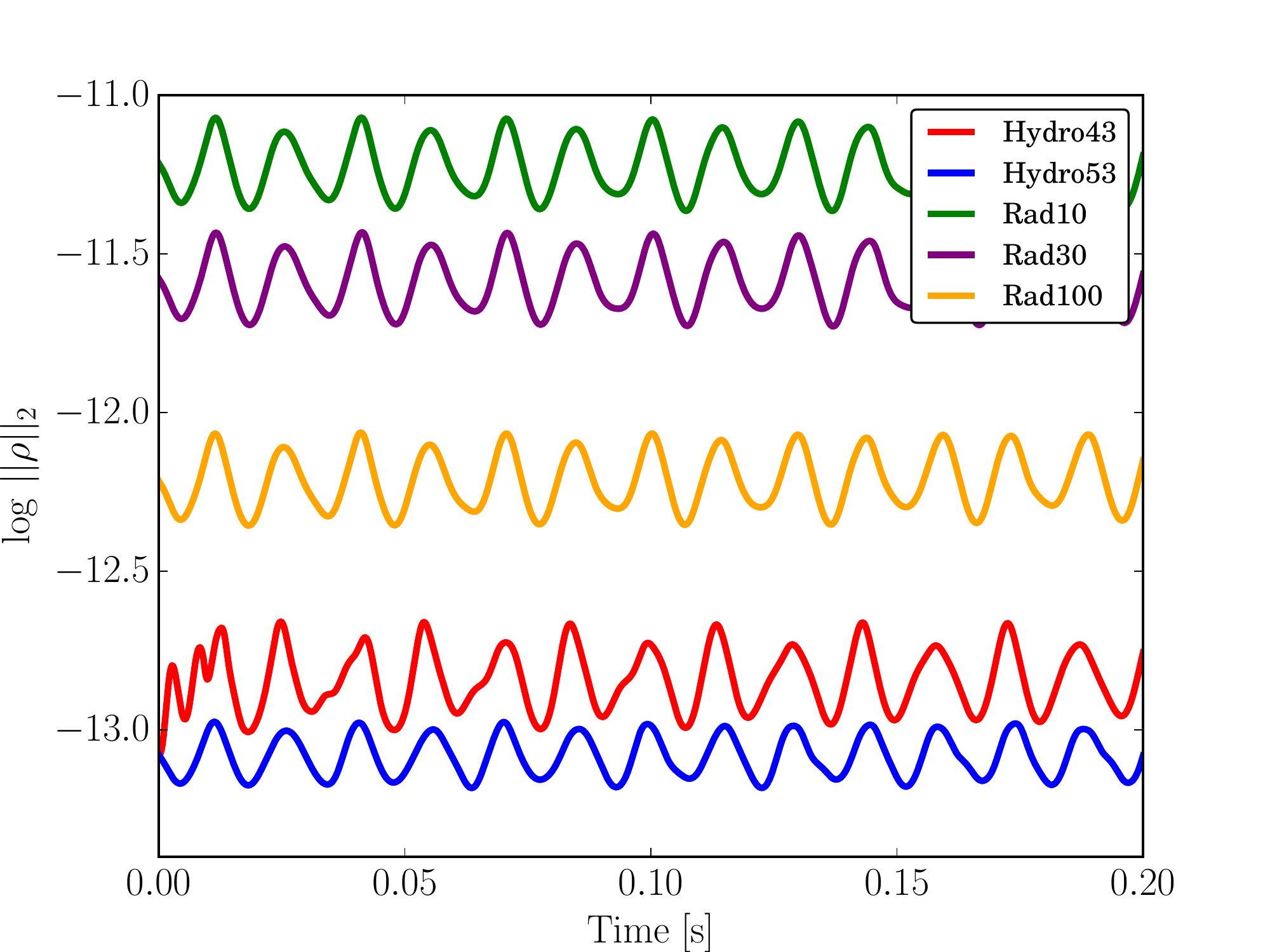}
	\caption{Evolution of the $L_2$ norm of the rest-mass density up to $t=0.2 \ {\rm s}$. 
	The values for the radiative models had been rescaled to be visible in approximately the same range of the hydrodynamic models.
	See Fig~\ref{fig:L2Rad400} for the same quantity in the model Rad400. 
	}
	\label{fig:L2All}
\end{figure}

\begin{figure}
	\centering
	\includegraphics[width=1.075\columnwidth]{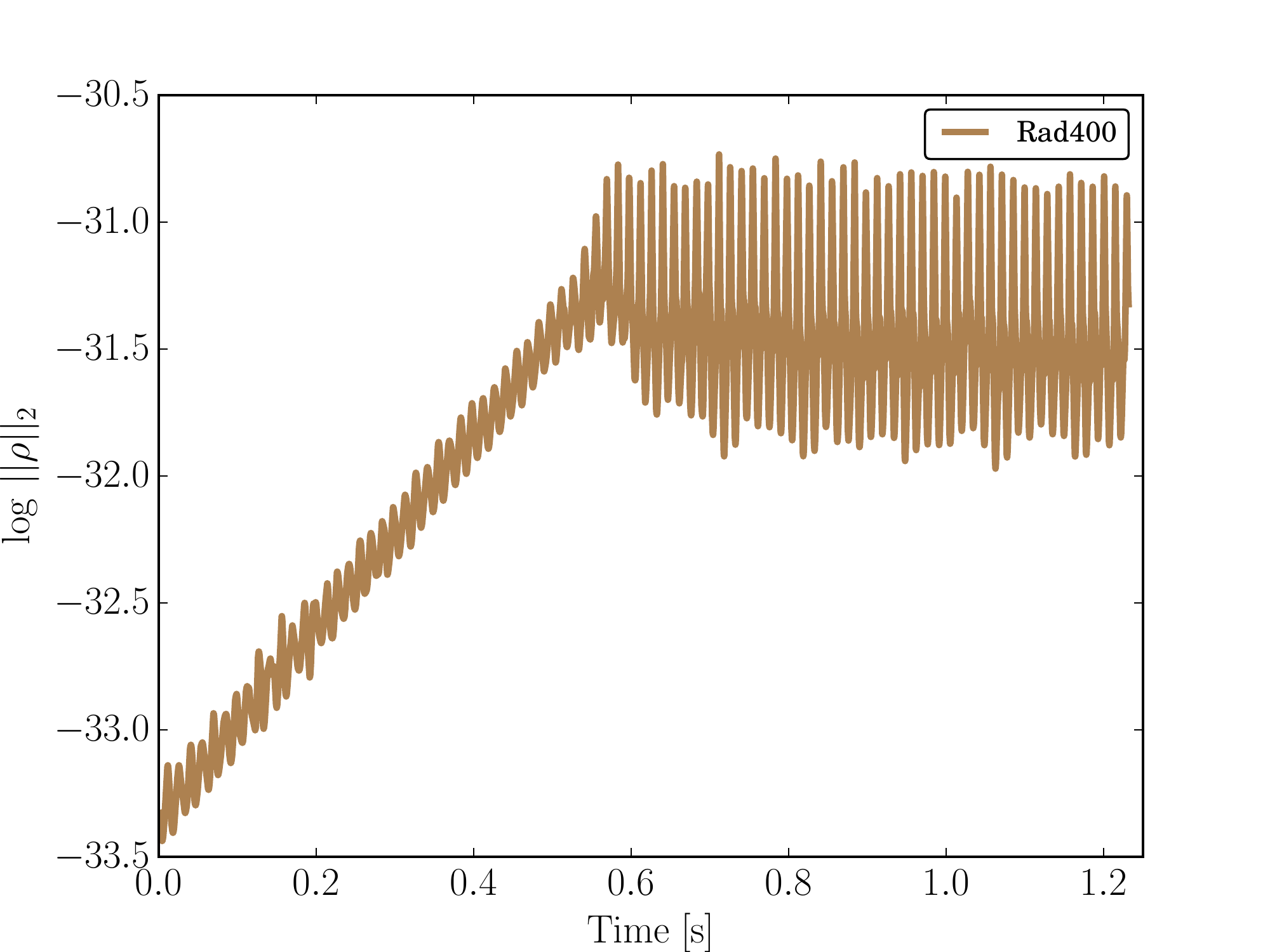}
	\caption{Evolution of the $L_2$ norm of the rest-mass density for the model
Rad400. 
}
	\label{fig:L2Rad400}
\end{figure}

However, by far the most convenient quantity to highlight the periodic response of the torus to the external perturbation is the $L_2$-norm of the rest-mass density, 
computed as, 
\begin{equation}
||\rho||_2=\sum_{ij} \rho_{ij}^2 dV_{ij}\,,
\end{equation}
where the summation extends all over the computational domain and
$\rho_{ij}$ and $dV_{ij}$ are the density and the volume of cell
\textit{ij}, respectively. 
We have reported this quantity in Fig.~\ref{fig:L2All}. 
It is interesting to note that, at least qualitatively, the amplitude of the oscillation, as well as its frequency, do not seem to change appreciably among the different models, nor they seem to depend on whether there is a radiation field or not. However, these comments will become quantitative in the discussion of Sect.~\ref{sec:spectral}.
The peculiar behaviour of the model Rad400 is represented in
Fig.~\ref{fig:L2Rad400}. As commented above, this model has a huge
initial value of the ratio $p_{\rm rad}/p_{\rm gas}$, but it is also
the least optically thick compared to the others, exhausting rapidly
its radiation content. As a result of the drastic size reduction, the
rest-mass density and the corresponding $L_2$-norm increase with time, until the torus enters a gas pressure dominated phase. This transition is well represented in Fig.~\ref{fig:L2Rad400}, where the $L_2$-norm of the rest-mass density settles into an oscillating plateau after $t\sim 0.52$\,s.


%

%================================================================
\subsection{Spectral analysis}
\label{sec:spectral}
\begin{figure}
	\centering
	\includegraphics[width=1.2\columnwidth,height=0.33\textheight]{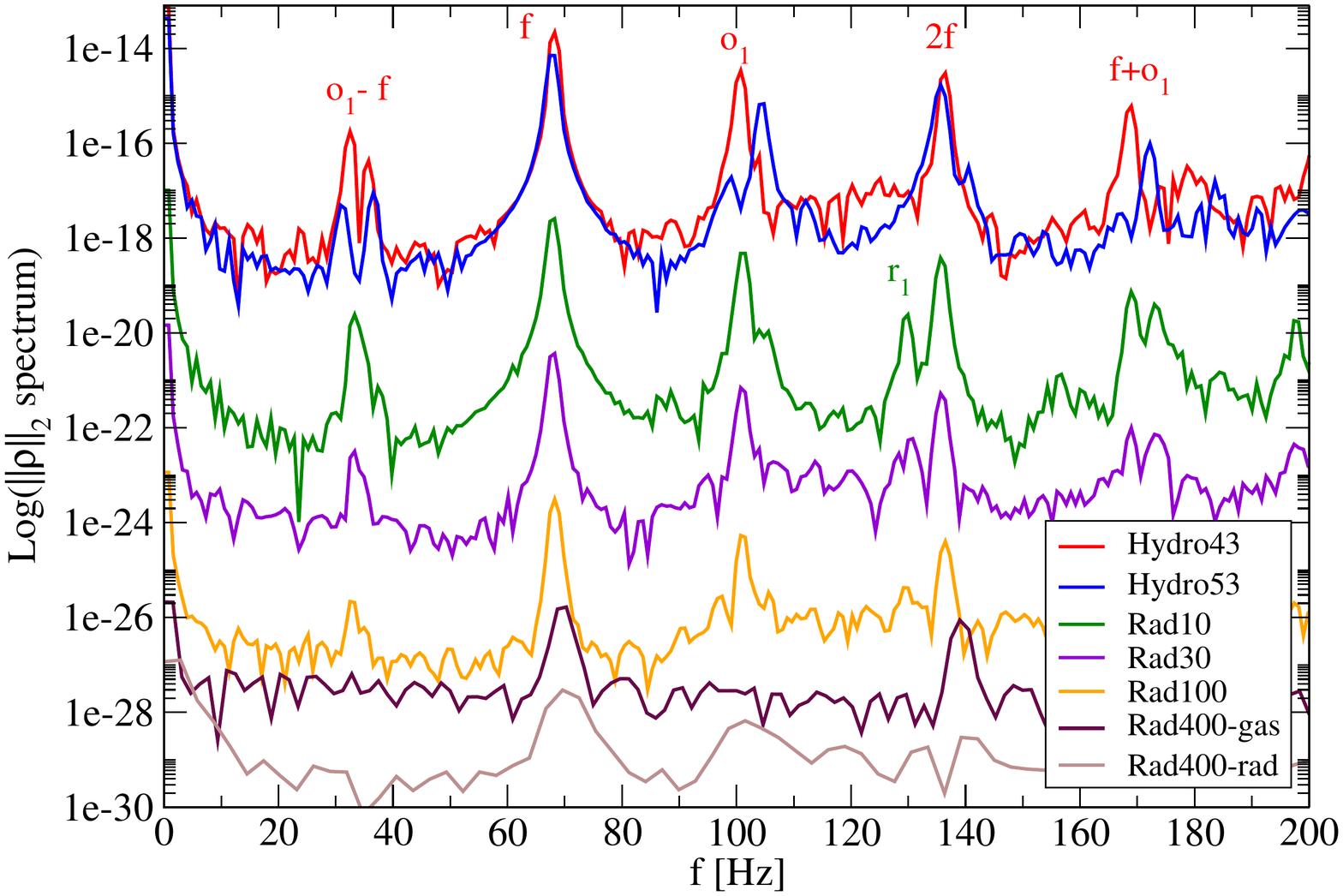}
	\caption{Spectrum of the $L_2$ norm of the rest-mass density. The fundamental mode, the overtones and their linear combinations are marked for the Hydro43 model (red curve).
	The units of the vertical axis are arbitrary.}
	\label{l2norm-spectra}
\end{figure}

It is interesting to investigate the effects of radiation field on the spectra 
of oscillating tori. To this extent, we have performed a Fourier analysis of the $L_2$-norm of the rest mass density within the torus, which
is reported in Fig.~\ref{l2norm-spectra}. Being a global quantity, the $L_2$-norm is particularly suitable for highlighting 
global modes of oscillations.
The first two spectra, marked in red and in blue, correspond to the purely hydrodynamic models, 
and, as usual, they serve us as a reference when comparing with the radiative models. In practice, 
two inertial-acoustic modes are detected, i.e. a fundamental mode ${\rm f}$ and an overtone ${\rm o_1}$, with a ratio ${\rm f/o_1}\sim1.47$.
In addition to them,
a sequence of linear combinations, such as $2{\rm f}$, ${\rm f+o_1}$, ${\rm o_1-f}$, are also found, which is a well known coupling effect in
physical systems governed by non-linear equations. All these results confirm what already found by \cite{Zanotti03,Zanotti05}
and are also in very good agreement with linear perturbative analysis performed by \cite{Rezzolla_qpo_03b}. Although it is not our aim 
to extend the linear perturbative analysis to radiation-hydrodynamics tori in equilibrium, it is interesting to note that, for our specific hydrodynamics models
with $\ell=3.8$, the value of the ratio ${\rm o_1/f}$ predicted by the linear analysis of \cite{Rezzolla_qpo_03b} is 1.47, 
which matches perfectly with that inferred from the red curve in Fig.~\ref{l2norm-spectra}.
There is however a detail which was not noticed in the previous investigations, namely the fact that, while the fundamental mode ${\rm f}$ does not depend on the polytropic index $\Gamma$,
the overtone ${\rm o_1}$ does depend on it, slightly increasing with increasing $\Gamma$. This may indicate that in the mode ${\rm o_1}$ 
the acoustic contribution is significant, while in the mode ${\rm f}$ the inertial contribution is dominant. We also recall that the frequency of the fundamental mode ${\rm f}$ tends to the epicyclic frequency at the radius of the maximum density in the limit of vanishing size tori~\citep{Rezzolla_qpo_03b}. 
In our models the epicyclic frequency at $r_c=8.35$ (the maximum density radius) is $f_{\rm epic}\sim 71$\,Hz, while the fundamental model ${\rm f}$ 
in Fig.~\ref{l2norm-spectra} is at $68$\,Hz.\footnote{These numbers apply for our specific torus model and assuming $M_{\rm BH}=10M_{\odot}$.}
Hence, the major contribution to ${\rm f}$ comes from the epicyclic term, and the dependence on $\Gamma$ is essentially lost.
We also note that the ${\rm f}$ and ${\rm o_1}$ modes found here may correspond to the so called {\em radial} and {\em plus} modes of slender tori, according to the terminology introduced by \cite{Blaes2006}.\footnote{See also the discussion in \cite{Mishra2015}.}

When migrating to the radiative models, the spectra maintain a good qualitative agreement with the purely hydrodynamic ones. In particular, all the modes ${\rm f}$, ${\rm o_1}$ and ${\rm 2f}$ are
still clearly visible. Apart from the model Rad400, which becomes gas-pressure dominated after $t\sim 0.5$\,s, all the others are radiation-pressure dominated all over the evolution. Consistently with that, their mode ${\rm o_1}$, which is sensible to the value of $\Gamma$, approaches the value obtained by the Hydro43 model with $\Gamma=4/3$.

Moreover, in the spectra of Rad10, Rad30 and Rad100 there are good indications for the presence of a genuinely new mode of oscillation, marked as $r_1$ for the model Rad10 at approximately
$r_1\sim6.39\times10^{-3}M_{\rm BH}^{-1}\sim129.8 M_{\rm BH}^{-1}\,\rm Hz $ 
and totally absent in the spectra of Hydro43 and Hydro53. Our claim that the mode  $r_1$ is due to the radiation field
could be rigorously confirmed after performing a linear perturbative
analysis of the relativistic radiation-hydrodynamics equations, which is however far beyond the scope of the present work.
The spectrum of the model Rad400 is much more noisy than the others and in practice only the fundamental mode ${\rm f}$ is clearly visible. This is
true both in the radiation-pressure phase of the evolution, with a corresponding spectrum marked as Rad400-rad in Fig.~\ref{l2norm-spectra}, and in the
gas-pressure phase, with a corresponding spectrum marked as Rad400-gas. This is not surprising, given that the dynamics of this model is so violent, causing the torus to collapse 
to a much thinner disc and no longer able to work as a cavity for the inertial-acoustic modes.

\begin{figure}
	\centering
	\includegraphics[width=1.2\columnwidth,height=0.33\textheight]{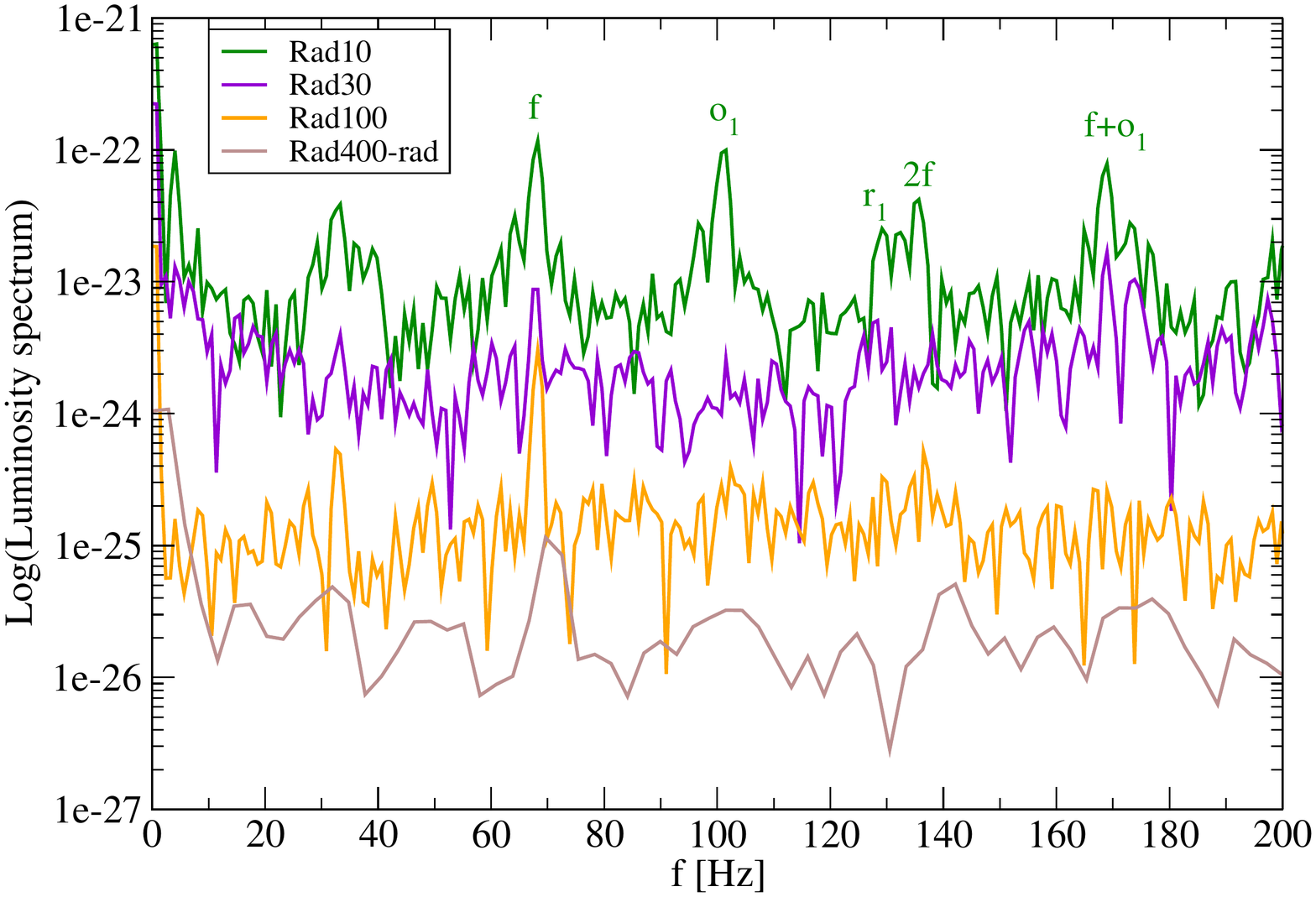}
	\caption{Spectrum of the luminosity. The units of the vertical axis are arbitrary.}
	\label{luminosity-spectra}
\end{figure}

For comparison, in Fig.~\ref{luminosity-spectra} we have reported also
the spectra of the luminosity (Fig.~\ref{fig:lum}), pertinent to the radiative models. 
Although these spectra reproduce the same kind of pattern described above, they are 
significantly more noisy than those of Fig.~\ref{l2norm-spectra}. In particular, the identification of the mode  $r_1$ is much more dubious. 
From one side, this is understandable, since the luminosity is a highly non-linear derived quantity. From another side, this is disappointing, since this is the quantity 
that would be more prone to potential observations, but from which we could not extract the richer information that is contained in Fig.~\ref{l2norm-spectra}.
Our results should also be compared with those of \cite{Schnittman06}, who computed the light curves from oscillating tori by using a ray-tracing code. 
In their case the power spectra of
the light curves are  equivalent to those obtained from the spectra of purely dynamic quantities. 
As we have shown, however, when the back reaction of radiation onto matter is properly taken into account through a time dependent radiation-hydrodynamic code, the 
imprint of the dynamics into the observable quantities is less transparent.

\section{Conclusions}
\label{s.discussion}
In this work we have studied the radiation-hydrodynamics evolution of relativistic tori 
that are subject to external perturbations. 
To this extent, we have performed high resolution two dimensional 
general relativistic simulations in the Schwarzschild metric
with the \texttt{KORAL} code, which can successfully
capture both the optically thick and the optically thin regime through the relativistic version of the M1-closure.
The evolution is followed for $\sim 165$ orbital periods.
Depending on the optical thickness of the initial configuration, the torus can either remain radiation pressure dominated, or migrate towards the gas pressure dominated
regime, after experiencing a rapid dynamical collapse. The periodic response induced by the external perturbation manifests in all the thermodynamics quantities, and in particular in the emitted luminosity, which in the most optically thick models can reach values as high as $L\sim 10^{38}-10^{39}\, {\rm erg \,s^{-1}}$. 

In addition to the standard inertial-acoustic oscillations already highlighted by \cite{Zanotti05} for purely hydrodynamic models, we have found a
new mode in the spectra of the $L_2$-norm of the rest-mass density,
with a clear peak at a frequency $129 M_{\rm BH}^{-1}\,\rm Hz$. This mode, which is absent 
in the purely hydrodynamic evolution, can be 
attributed to the radiation with a high degree of fidelity. These results suggest that the interpretation of kHz-QPOs in the X-ray spectra of Low Mass X-ray binaries
in terms of inertial-acoustic oscillations \citep{Rezzolla_qpo_03a} must necessarily be extended to take into account the role of a radiation field.

Finally, the power spectra obtained from the light curves are substantially more noisy than those extracted from the dynamic quantities, 
which should sound as a caveat for those cases in which the interaction of radiation and matter is not properly taken into account.
\section{Acknowledgements}
G.P.M. acknowledges support by Polish NCN grant 2013/09/B/ST9/00060.
O.Z. acknowledges support by the
European Research Council (ERC) under the
European Union's Seventh Framework Programme (FP7/2007-2013), 
ERC Grant agreement no. 278267.
A.S.\ acknowledges support by NASA through Einstein Postdoctoral Fellowship PF4-150126
awarded by the Chandra X-ray Center, which is operated by the
Smithsonian Astrophysical Observatory for NASA under contract NAS8-03060. 

We would also like to acknowledge PRACE for awarding access to the SuperMUC 
supercomputer based in Munich, Germany at the Leibniz Rechenzentrum (LRZ).

\bibliographystyle{mn2e}
\bibliography{references}

\begin{thebibliography}{}

\bibitem[\protect\citeauthoryear{{Abramowicz}, {Jaroszynski} \&
  {Sikora}}{{Abramowicz} et~al.}{1978}]{Abramowicz78}
{Abramowicz} M.,  {Jaroszynski} M.,    {Sikora} M.,  1978, Astron. Astrophys.,
  63, 221

\bibitem[\protect\citeauthoryear{Abramowicz \& Fragile}{Abramowicz \&
  Fragile}{2013}]{Abramowicz2011}
Abramowicz M.~A.,  Fragile P.~C.,  2013, Living Reviews in Relativity, 16

\bibitem[\protect\citeauthoryear{{Abramowicz}, {Karas}, {Kluzniak}, {Lee} \&
  {Rebusco}}{{Abramowicz} et~al.}{2003}]{Abramowicz2003}
{Abramowicz} M.~A.,  {Karas} V.,  {Kluzniak} W.,  {Lee} W.~H.,    {Rebusco} P.,
   2003, Pub. Astron. Soc. Japan, 55, 467

\bibitem[\protect\citeauthoryear{{Bakala}, {Goluchov{\'a}}, {T{\"o}r{\"o}k},
  {{\v S}r{\'a}mkov{\'a}}, {Abramowicz}, {Vincent} \& {Mazur}}{{Bakala}
  et~al.}{2015}]{Bakala2015}
{Bakala} P.,  {Goluchov{\'a}} K.,  {T{\"o}r{\"o}k} G.,  {{\v S}r{\'a}mkov{\'a}}
  E.,  {Abramowicz} M.~A.,  {Vincent} F.~H.,    {Mazur} G.~P.,  2015, Astron.
  Astrophys., 581, A35

\bibitem[\protect\citeauthoryear{{Blaes}, {Arras} \& {Fragile}}{{Blaes}
  et~al.}{2006}]{Blaes2006}
{Blaes} O.~M.,  {Arras} P.,    {Fragile} P.~C.,  2006, Mon. Not. R. Astron.
  Soc., 369, 1235

\bibitem[\protect\citeauthoryear{{Blaes}, {{\v S}r{\'a}mkov{\'a}},
  {Abramowicz}, {Klu{\'z}niak} \& {Torkelsson}}{{Blaes}
  et~al.}{2007}]{Blaes2007}
{Blaes} O.~M.,  {{\v S}r{\'a}mkov{\'a}} E.,  {Abramowicz} M.~A.,
  {Klu{\'z}niak} W.,    {Torkelsson} U.,  2007, Astrophys. J., 665, 642

\bibitem[\protect\citeauthoryear{{Farris}, {Li}, {Liu} \& {Shapiro}}{{Farris}
  et~al.}{2008}]{Farris08}
{Farris} B.~D.,  {Li} T.~K.,  {Liu} Y.~T.,    {Shapiro} S.~L.,  2008, Phys.
  Rev. D, 78, 024023

\bibitem[\protect\citeauthoryear{{Fishbone} \& {Moncrief}}{{Fishbone} \&
  {Moncrief}}{1976}]{Fishbone76}
{Fishbone} L.~G.,  {Moncrief} V.,  1976, Astrophys. J., 207, 962

\bibitem[\protect\citeauthoryear{Font \& Daigne}{Font \&
  Daigne}{2002}]{Font02a}
Font J.~A.,  Daigne F.,  2002, Mon. Not. R. Astron. Soc., 334, 383

\bibitem[\protect\citeauthoryear{{Fragile}, {Gillespie}, {Monahan}, {Rodriguez}
  \& {Anninos}}{{Fragile} et~al.}{2012}]{Fragile2012}
{Fragile} P.~C.,  {Gillespie} A.,  {Monahan} T.,  {Rodriguez} M.,    {Anninos}
  P.,  2012, ArXiv e-prints

\bibitem[\protect\citeauthoryear{{Kozlowski}, {Jaroszynski} \&
  {Abramowicz}}{{Kozlowski} et~al.}{1978}]{Kozlowski1978}
{Kozlowski} M.,  {Jaroszynski} M.,    {Abramowicz} M.~A.,  1978, Astron. and
  Astrophys., 63, 209

\bibitem[\protect\citeauthoryear{{Levermore}}{{Levermore}}{1984}]{levermore+84}
{Levermore} C.~D.,  1984, \jqsrt, 31, 149

\bibitem[\protect\citeauthoryear{{Mazur}, {Vincent}, {Johansson}, {{\v
  S}ramkov{\'a}}, {T{\"o}r{\"o}k}, {Bakala} \& {Abramowicz}}{{Mazur}
  et~al.}{2013}]{Mazur2013}
{Mazur} G.~P.,  {Vincent} F.~H.,  {Johansson} M.,  {{\v S}ramkov{\'a}} E.,
  {T{\"o}r{\"o}k} G.,  {Bakala} P.,    {Abramowicz} M.~A.,  2013, Astron.
  Astrophys., 554, A57

\bibitem[\protect\citeauthoryear{{McKinney}, {Tchekhovskoy}, {Sadowski} \&
  {Narayan}}{{McKinney} et~al.}{2013}]{McKinney2013b}
{McKinney} J.~C.,  {Tchekhovskoy} A.,  {Sadowski} A.,    {Narayan} R.,  2013,
  ArXiv e-prints

\bibitem[\protect\citeauthoryear{Michel}{Michel}{1972}]{michel72}
Michel F.~C.,  1972, Astrophys. Spa. Sci., 15, 153

\bibitem[\protect\citeauthoryear{{Mishra}, {Vincent}, {Manousakis}, {Fragile},
  {Paumard} \& {Klu{\'z}niak}}{{Mishra} et~al.}{2015}]{Mishra2015}
{Mishra} B.,  {Vincent} F.~H.,  {Manousakis} A.,  {Fragile} P.~C.,  {Paumard}
  T.,    {Klu{\'z}niak} W.,  2015, ArXiv e-prints

\bibitem[\protect\citeauthoryear{{Montero} \& {Zanotti}}{{Montero} \&
  {Zanotti}}{2012}]{Montero2012}
{Montero} P.~J.,  {Zanotti} O.,  2012, Mon. Not. R. Astron. Soc., 419, 1507

\bibitem[\protect\citeauthoryear{{Novikov} \& {Thorne}}{{Novikov} \&
  {Thorne}}{1973}]{Novikov:1973}
{Novikov} I.~D.,  {Thorne} K.~S.,  1973, in Black Holes (Les Astres Occlus)
  {Astrophysics of black holes.}.
pp 343--450

\bibitem[\protect\citeauthoryear{{Remillard} \& {McClintock}}{{Remillard} \&
  {McClintock}}{2006}]{Remillard2006}
{Remillard} R.~A.,  {McClintock} J.~E.,  2006, Ann. Rev. Astron. Astroph., 44,
  49

\bibitem[\protect\citeauthoryear{{Rezzolla}, {Yoshida}, {Maccarone} \&
  {Zanotti}}{{Rezzolla} et~al.}{2003}]{Rezzolla_qpo_03a}
{Rezzolla} L.,  {Yoshida} S.,  {Maccarone} T.~J.,    {Zanotti} O.,  2003, Mon.
  Not. R. Astron. Soc., 344, L37

\bibitem[\protect\citeauthoryear{{Rezzolla}, {Yoshida} \& {Zanotti}}{{Rezzolla}
  et~al.}{2003}]{Rezzolla_qpo_03b}
{Rezzolla} L.,  {Yoshida} S.,    {Zanotti} O.,  2003, Mon. Not. R. Astron.
  Soc., 344, 978

\bibitem[\protect\citeauthoryear{{Roedig}, {Zanotti} \& {Alic}}{{Roedig}
  et~al.}{2012}]{Roedig2012}
{Roedig} C.,  {Zanotti} O.,    {Alic} D.,  2012, Mon. Not. R. Astron. Soc.,
  426, 1613

\bibitem[\protect\citeauthoryear{{Sadowski} \& {Narayan}}{{Sadowski} \&
  {Narayan}}{2015}]{sadowski+compt}
{Sadowski} A.,  {Narayan} R.,  2015, ArXiv e-prints

\bibitem[\protect\citeauthoryear{{S{\c a}dowski}, {Narayan}, {McKinney} \&
  {Tchekhovskoy}}{{S{\c a}dowski} et~al.}{2014}]{Sadowski2014}
{S{\c a}dowski} A.,  {Narayan} R.,  {McKinney} J.~C.,    {Tchekhovskoy} A.,
  2014, \mnras, 439, 503

\bibitem[\protect\citeauthoryear{{S{\c a}dowski}, {Narayan}, {Tchekhovskoy} \&
  {Zhu}}{{S{\c a}dowski} et~al.}{2013}]{Sadowski2013}
{S{\c a}dowski} A.,  {Narayan} R.,  {Tchekhovskoy} A.,    {Zhu} Y.,  2013, Mon.
  Not. R. Astron. Soc., 429, 3533

\bibitem[\protect\citeauthoryear{{Schnittman} \& {Rezzolla}}{{Schnittman} \&
  {Rezzolla}}{2006}]{Schnittman06}
{Schnittman} J.~D.,  {Rezzolla} L.,  2006, Astrophys. J., 637, L113

\bibitem[\protect\citeauthoryear{{Shakura} \& {Sunyaev}}{{Shakura} \&
  {Sunyaev}}{1973}]{Shakura1973}
{Shakura} N.~I.,  {Sunyaev} R.~A.,  1973, Astron. Astrophys., 24, 337

\bibitem[\protect\citeauthoryear{{Takahashi} \& {Ohsuga}}{{Takahashi} \&
  {Ohsuga}}{2013}]{Takahashi2013}
{Takahashi} H.~R.,  {Ohsuga} K.,  2013, Astrophys. J., 772, 127

\bibitem[\protect\citeauthoryear{{Vincent}, {Mazur}, {Straub}, {Abramowicz},
  {Klu{\'z}niak}, {T{\"o}r{\"o}k} \& {Bakala}}{{Vincent}
  et~al.}{2014}]{Vincent2014}
{Vincent} F.~H.,  {Mazur} G.~P.,  {Straub} O.,  {Abramowicz} M.~A.,
  {Klu{\'z}niak} W.,  {T{\"o}r{\"o}k} G.,    {Bakala} P.,  2014, Astron.
  Astrophys., 563, A109

\bibitem[\protect\citeauthoryear{{Zanotti}}{{Zanotti}}{2014}]{Zanotti2014}
{Zanotti} O.,  2014, Astronomy and Astrophysics, 563, A17

\bibitem[\protect\citeauthoryear{{Zanotti}, {Font}, {Rezzolla} \&
  {Montero}}{{Zanotti} et~al.}{2005}]{Zanotti05}
{Zanotti} O.,  {Font} J.~A.,  {Rezzolla} L.,    {Montero} P.~J.,  2005, Mon.
  Not. R. Astron. Soc., 356, 1371

\bibitem[\protect\citeauthoryear{Zanotti, Rezzolla \& Font}{Zanotti
  et~al.}{2003}]{Zanotti03}
Zanotti O.,  Rezzolla L.,    Font J.~A.,  2003, Mon. Not. Roy. Soc., 341, 832

\bibitem[\protect\citeauthoryear{{Zanotti}, {Roedig}, {Rezzolla} \& {Del
  Zanna}}{{Zanotti} et~al.}{2011}]{Zanotti2011}
{Zanotti} O.,  {Roedig} C.,  {Rezzolla} L.,    {Del Zanna} L.,  2011, Mon. Not.
  R. Astron. Soc., 417, 2899

\end{thebibliography}

%\begin{thebibliography}
%\bibitem[Abramowicz et al.(1988)]{abra88} Abramowicz, M.~A., Czerny,
%  B., Lasota, J.~P., \& Szuszkiewicz, E.\ 1988, \apj, 332, 646
%\bibitem[Abramowicz et al.(1978)]{abra78} Abramowicz, M., Jaroszynski, M., \& Sikora, M.\ 1978, \aap, 63, 221 
%\bibitem[Michel(1972)]{michel72} Michel, F.~C.\ 1972, \apss, 15, 153 
%\bibitem[Zanotti et al.(2003)]{zanotti03} Zanotti, O., Rezzolla, 
%L., \& Font, J.~A.\ 2003, \mnras, 341, 832 
%\end{thebibliography}

\end{document}